\documentclass[sn-basic, Numbered, iicol]{sn-jnl}

\usepackage{graphicx}
\usepackage{multirow}
\usepackage{amsmath,amssymb,amsfonts}
\usepackage{amsthm}
\usepackage{mathrsfs}
\usepackage[title]{appendix}
\usepackage{xcolor}
\usepackage{textcomp}
\usepackage{manyfoot}
\usepackage{booktabs}
\usepackage{hyperref}
\usepackage{algorithm}
\usepackage{algorithmicx}
\usepackage{algpseudocode}
\usepackage{listings}
\usepackage{float}
\usepackage{placeins}
\usepackage{svg}
\usepackage{xr-hyper} 
\theoremstyle{thmstyleone}

\theoremstyle{thmstyletwo}

\theoremstyle{thmstylethree}

\usepackage{amsmath,amsfonts,bm}

\def\eqref#1{equation~\ref{#1}}

\def\1{\bm{1}}

\DeclareMathAlphabet{\mathsfit}{\encodingdefault}{\sfdefault}{m}{sl}
\SetMathAlphabet{\mathsfit}{bold}{\encodingdefault}{\sfdefault}{bx}{n}

\newcommand{\norm}[1]{\left| #1 \right|}
\newcommand\dd{\mathop{}\!\mathrm{d}}

\DeclareMathOperator*{\argmin}{arg\,min}

\newcommand{\name}{LAVA }
\newcommand{\sptable}[1]{Supplementary Table #1}

\newcommand{\spnote}[1]{Supplementary Note #1}

\begin{document}

\title{Neural Scaling Laws Surpass Chemical Accuracy for the Many-Electron Schrödinger Equation}

\author[1, 2]{\fnm{Du} \sur{Jiang}}
\equalcont{These authors contributed equally to this work.}

\author*[1]{\fnm{Xuelan} \sur{Wen}}
\email{wxl@bytedance.com}
\equalcont{These authors contributed equally to this work.}

\author[1]{\fnm{Yixiao} \sur{Chen}}

\author[1, 2]{\fnm{Ruichen} \sur{Li}}

\author[1, 2]{\fnm{Weizhong} \sur{Fu}}

\author[1]{\fnm{Hung Q.} \sur{Pham}}

\author*[2]{\fnm{Ji} \sur{Chen}}
\email{ji.chen@pku.edu.cn}

\author*[2]{\fnm{Di} \sur{He}}
\email{dihe@pku.edu.cn}

\author*[3]{\fnm{William A.} \sur{Goddard III}}
\email{wag@caltech.edu}

\author*[2]{\fnm{Liwei} \sur{Wang}}
\email{wanglw@pku.edu.cn}

\author*[1]{\fnm{Weiluo} \sur{Ren}}
\email{renweiluo@bytedance.com}

\affil*[1]{ByteDance Seed}
\affil*[2]{Peking University}
\affil*[3]{California Institute of Technology}

\abstract{

We demonstrate, for the first time, that neural scaling laws can deliver near-exact solutions to the many-electron Schrödinger equation across a broad range of realistic molecules.
This progress is enabled by the Lookahead Variational Algorithm (LAVA), an effective optimization scheme that systematically translates increased model size and computational resources into greatly improved energy accuracy for neural network wavefunctions.
Across all tested cases, including benzene, the absolute energy error exhibits a systematic power-law decay with respect to model capacity and computation resources.
The resulting energies not only surpass the 1 kcal/mol “chemical-accuracy” threshold but also achieve 1 kJ/mol subchemical accuracy.
Beyond energies, the scaled-up neural network also yields better wavefunctions with improved physical symmetries, alongside accurate electron densities, dipole moments, and other important properties.
Our approach offers a promising way forward to addressing many long-standing challenges in quantum chemistry.
For instance, we improve energetic properties for systems such as the potential energy curve of nitrogen dimer as dissociation is approached and the cyclobutadiene automerization reaction barrier, producing definitive benchmarks, particularly in regimes where experimental data are sparse or highly uncertain.
We also shed light on the decades-old puzzle of the cyclic ozone stability with highly accurate calculations for the cyclic-to-open ozone barrier.
These results provide near-exact reference calculations with unprecedented accuracy, universal reliability and practical applicability, establishing a foundation for AI-driven quantum chemistry.
}

\maketitle

\section{Main}\label{sec1}
The many-electron Schrödinger equation lies at the foundation of quantum chemistry and condensed matter physics, providing a first-principles framework for understanding the quantum nature of the physical world.
Despite its central importance, no general-purpose method has come close to solving it exactly for realistic systems.
Instead, practical quantum chemistry has long relied on the cancellation of large and often uncontrolled errors to reach the so-called “chemical accuracy”--typically defined as energy errors within 1 kcal/mol for relevant energy differences.
This scheme, however, comes with several significant limitations, including poor performance in predicting observables beyond relative energies and a lack of systematic error control.
For instance, the widely used Density Functional Theory (DFT) can predict qualitatively incorrect electron densities~\cite{Medvedev2017}, undermining its reliability for density-derived properties, such as dipole moments and polarizabilities.
Correlated wavefunction methods, on the other hand, depend on error cancellation due to steep computational scaling and slow basis-set convergence~\cite{bak2000coupled, peterson2012chemical}, compromising their reliability in complex problems.
These limitations underscore the need for a more accurate and reliable solution that delivers cancellation-free, high-accuracy energies and other observables for realistic systems.

Neural network-based quantum Monte Carlo (NNQMC) has emerged as a promising \textit{ab initio} wavefunction theory to this challenge~\cite{carleo2017solving, hermann_deep-neural-network_2020,ferminet}.
Unlike other machine learning approaches in quantum chemistry that rely on precomputed labeled data (e.g., DFT energies)~\cite{li2022deep,chandrasekaran2019solving}, NNQMC obtains the target quantum state directly through unsupervised optimization, without requiring any reference data.
In particular, the full many-body wavefunction is modeled with highly expressive neural networks, providing access to accurate total energies, high-quality wavefunctions, and derived observables including electron and spin densities.
Recent progress has improved energy accuracy as well as computational efficiency, highlighting its potential as a next-generation quantum chemistry framework~\cite{li_computational_2024, li_spin-symmetry-enforced_2024, Pfau2024,  scherbela2025accurate, fu2025local, foster2025ab}.
Nonetheless, NNQMC has not yet come meaningfully close to the exact solutions of the Schrödinger equation as molecular systems increase in size and complexity.
This is partly because default-sized neural networks lack sufficient representational capacity for larger systems, while simply increasing the network size rarely leads to proportional improvements in accuracy due to optimization challenges.

In this work, we train neural‑network wavefunctions that, for the first time, deliver near‑exact solutions to the many‑electron Schrödinger equation for realistic systems with up to 12 atoms, achieving accuracy on par with experimental uncertainty.
The resulting absolute energies surpass traditional ``chemical accuracy'' to approach the 1 kJ/mol regime, thereby enabling definitive relative energies without relying on error cancellation.
Additionally, these solutions provide accurate many-body wavefunctions, which in turn produce benchmark-quality physical observables including electron density and dipole moments.
This capability is made possible by two key innovations.
First, we introduce the Lookahead Variational Algorithm (LAVA), an optimization scheme that combines variational and projective frameworks,  offering better performance over existing schemes in NNQMC, such as Variational Monte Carlo (VMC) with stochastic reconfiguration~\cite{PhysRevB.64.024512} and Wasserstein Quantum Monte Carlo (WQMC)~\cite{neklyudov2023wasserstein}.
Second, we present the first systematic study of neural scaling laws~\cite{kaplan2020scaling} in \textit{ab initio} quantum chemistry, showing that total energy errors decrease systematically and predictably by simply scaling up neural network model capacity and computational resources. 
Together, these advances position our approach uniquely within quantum chemistry, offering near-exact total energies and wavefunctions at the full configuration interaction (FCI) accuracy and complete basis set (CBS) limit at the same time.

Through this approach, we address several long-standing challenges in quantum chemistry, demonstrating its universal accuracy, flexibility, and ability to provide definitive benchmarks.
Firstly, we establish a high-quality benchmark for cyclobutadiene's transition barrier, aligned with refined experimental data as well as the best estimates from coupled cluster (CC) and configuration interaction (CI) methods.
Secondly, we study the potential energy curve (PEC) of diatomic molecules, which governs the vibrational levels critical for astrophysical models of planetary and stellar atmospheres~\cite{tennyson2016exomol}. 
In particular, we focus on nitrogen dimer and present a new PEC benchmark that surpasses previous experiment-based references in both accuracy and reliability.
Thirdly, we reassess the metastability of the cyclic ozone, helping to provide a definitive answer to a long-standing controversy~\cite{burton1977theoretical}.
Collectively, these findings confirm our approach as a new standard of accuracy for predictive quantum chemistry.

\section{Results}\label{sec2}

\subsection{Scale Toward Exact Solutions with LAVA}\label{subsec21}

\begin{figure*}[t]
    \centering 
    \includegraphics[width=\textwidth]{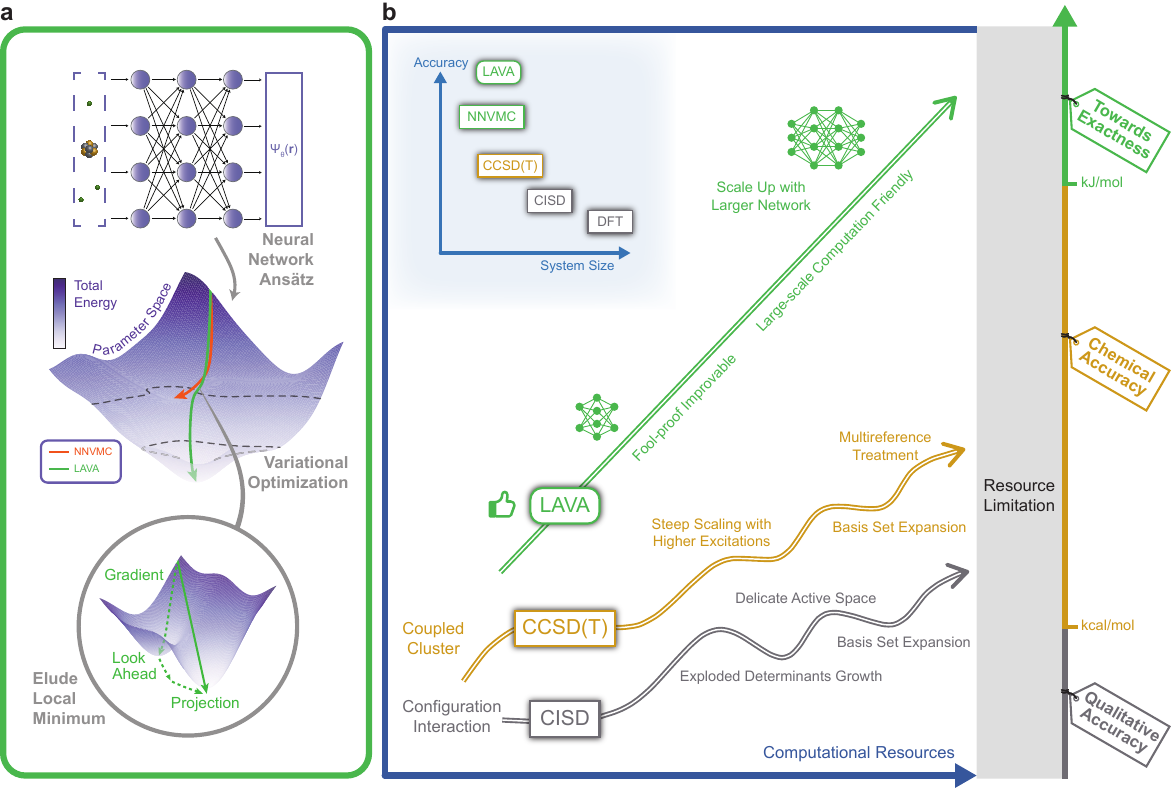}
    \caption{\small {\bf Scaling Up LAVA-Optimized Neural Network Wavefunctions Toward Exactness.} 
    \textbf{a.} Upper: The many-electron wavefunction is modeled by a massively parameterized neural network and optimized with LAVA or neural network-based variational Monte Carlo (NNVMC).
    Middle: A conceptual energy landscape illustrates LAVA’s better convergence behavior throughout the training process, compared to NNVMC, which only considers energy-based losses.
    Lower: LAVA combines both the gradient of an energy-based loss and a projection-derived direction via the Lookahead algorithm. 
    \textbf{b.} Main panel:  By scaling up the network size and computation resources, LAVA achieves systematic and fool-proof improvements in accuracy, surpassing chemical accuracy and approaching exact solutions.
    In contrast, traditional computational methods, such as coupled cluster and configuration interaction, suffer from steep computational scaling and resource bottlenecks, due to various issues such as basis set limitations, exploded determinant growth, and steep scaling with higher excitations.
     Inset: A schematic plot of different quantum chemistry methods in terms of accuracy and system size. 
    Notably, LAVA is able to achieve more accurate energy results than previous NNVMC works with the same neural network size.
    }
    \label{fig:overview}
\end{figure*}

\begin{figure*}[htp]
    \centering
    \includegraphics[width=1.0\textwidth]{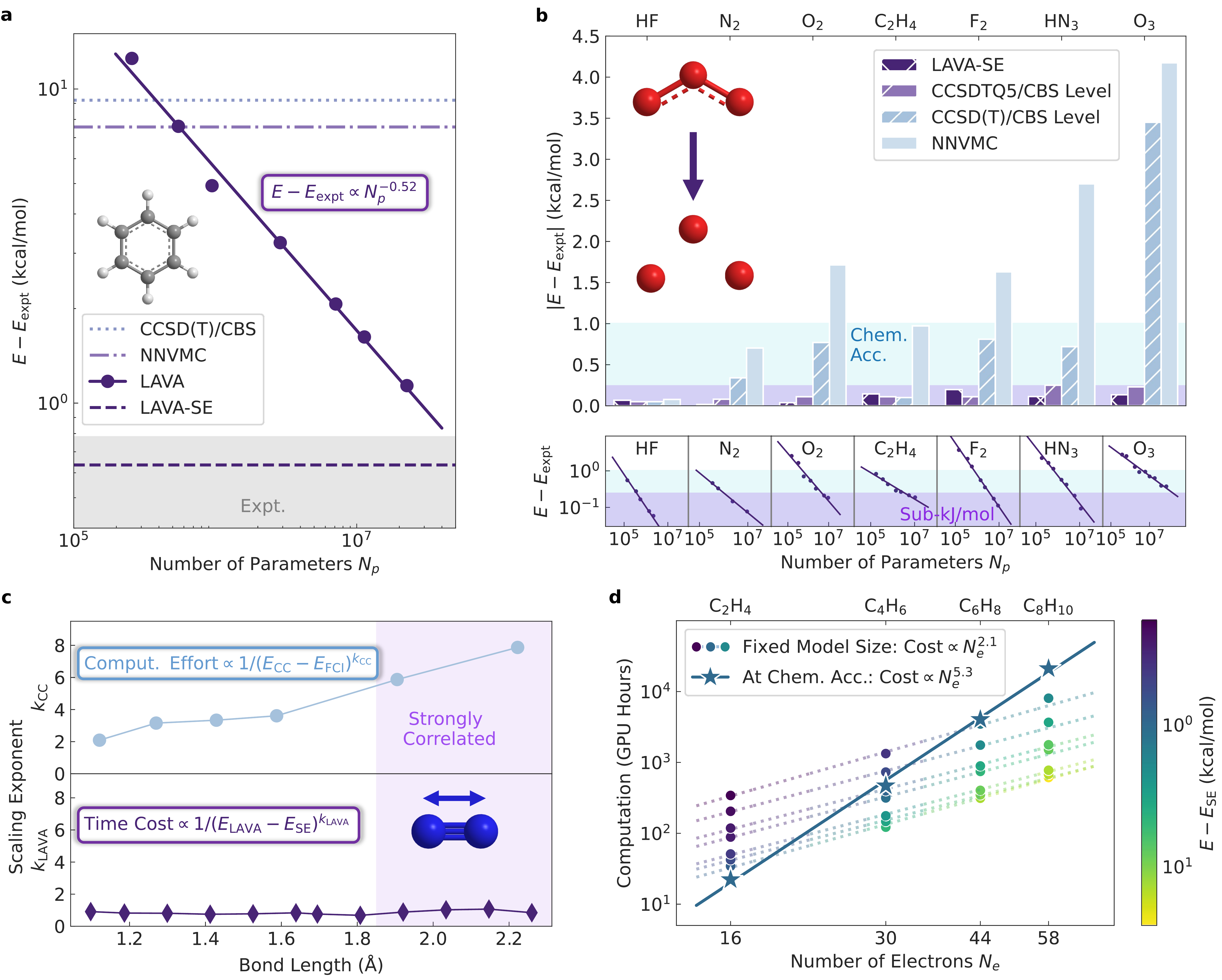}
    \caption{\small {\bf Neural scaling laws enable breakthrough performance.} 
    \textbf{a.} LAVA delivers near-exact ground state energy for benzene as neural networks scale up.
    The neural scaling law here is described by $E-E_{\mathrm{expt}}\propto N_p^{-0.52}$, where $E$ denotes LAVA energies, $E_{\mathrm{expt}}$ denotes experiment-derived reference~\cite{parthiban_fully_2001}, $N_p$ is the number of parameters in the neural networks.
    We also include energies from NNVMC~\cite{li_computational_2024} and CCSD(T)/CBS~\cite{Ren2023} for comparison.
    \textbf{b.} Upper: Absolute energy errors of LAVA-SE, CCSDTQ5/CBS level W4-theory~\cite{karton_w4-11_2011}, CCSD(T)/CBS level W2-theory~\cite{karton_w4-11_2011}, and NNVMC.
    LAVA-SE is more accurate than 1 kJ/mol thresholds (purple shade), aligning closely with highly accurate W4 theory.
    Experimental references are based on atomization energies from ATcT (Active Thermochemical Tables)~\cite{karton_w4-11_2011} and absolute atom energies from \citet{PhysRevA.47.3649}. Specifically, for atom $\text{F}$, we use our variationally lower result since it is more reliable. Details are in \spnote{4.1}.
    NNVMC uses LapNet~\cite{li_computational_2024} with the default model size. 
    Lower: Neural scaling laws of LAVA on the same molecules, showing error reduction with model size in power-law trends.
    See \spnote{2} for details.
    \textbf{c.} Cost scaling exponents for N$_2$ as LAVA increases model size and CC methods increase excitation orders. For CC, data from \citet{chan_state---art_2004} shows that CC's computational effort scales as a power-law trend relative to $1/(E_{\mathrm{CC}}-E_{\mathrm{FCI}})$ under cc-pVDZ basis set. For LAVA, the total GPU time cost scales as a power-law relative to $1/(E_{\mathrm{LAVA}}-E_{\mathrm{SE}})$, where $E_{\mathrm{SE}}$ denotes the LAVA-SE result.
    LAVA maintains a nearly constant exponent along the $N_2$ dissociation curve, while CC's performance deteriorates in the strongly correlated region.
    \textbf{d.} LAVA GPU hours as the number of electrons $N_e$ increases, with increasing model size for better accuracy.
    With model size fixed, runtime scales as $N_e^{2.1}$ (dotted lines).
    To ensure chemical accuracy, runtime scales as $N_e^{5.3}$ (the solid line).
    }
    \label{fig:benchmark}
\end{figure*}

In this section, we present a systematic study of neural scaling laws that achieve beyond-chemical-accuracy solutions to the Schr\"odinger equation. 
We demonstrate that neural network-based solutions systematically approach exact results with increasing model capacity and training compute.
Notably, the convergence curve of absolute energy error follows a robust power-law decay, and the same convergence behavior is consistently observed across diverse molecular systems.

Although scaling laws have reshaped various domains of Artificial Intelligence~\cite{doi:10.1073/pnas.2311878121,hestness2017deeplearningscalingpredictable,henighan2020scalinglawsautoregressivegenerative,brown2020languagemodelsfewshotlearners,Kaplan2020ScalingLF,hoffmann2022trainingcomputeoptimallargelanguage,Zhai_2022_CVPR,gordon-etal-2021-data}, their application within quantum science remains underexplored~\cite{qxc3-bkc7}.
More critically, the benefit of scaling up neural‑network wavefunctions appears limited: energy improvement tends to saturate well before reaching exactness, hinting at an inability to fully exploit the capacity of large neural networks.
To overcome this bottleneck, we introduce LAVA, an improved optimization framework for neural network wavefunctions that combines variational Monte Carlo updates with a projective step inspired by imaginary time evolution (Fig.~\ref{fig:overview}a). 
This two-step procedure is effective for eluding local minima during the neural network training process, which is crucial for achieving asymptotic exactness as the neural network ansatz scales up.
In practice, LAVA significantly improves stability during the training process and accuracy at the end of training, resulting in better wavefunctions and energies.
See Section~\ref{sec:method_lava} and \spnote{1.1} for algorithmic details.

LAVA scales predictably and robustly to deliver accurate solutions beyond chemical accuracy.
In pursuing systematic improvement towards exactness, our approach offers several key advantages.
Most notably, LAVA requires little heuristic tuning or chemical intuition, making it effectively a ``fool-proof'' process.
Moreover, LAVA avoids the prohibitive scaling with excitation order inherent to traditional methods such as coupled cluster theory, offering a significantly more efficient route to high accuracy.
These advantages are sketched in the main panel of Fig.~\ref{fig:overview}b.

To demonstrate the effectiveness of neural scaling laws, we provide a range of quantitative evidence, as illustrated in Fig.~\ref{fig:benchmark}. 
We examine the absolute energy of representative organic molecules, for which highly accurate experimental benchmarks (via total atomization energy) and theoretical references (W4 theory) are available.
Our energy estimates not only surpass chemical accuracy but also fall within experimental uncertainty, a more stringent criterion for accuracy.
Fig.~\ref{fig:benchmark}a illustrates how neural scaling laws enable such accuracy, using benzene as a representative example.
Specifically, the energy error decays following a power-law relation as the number of parameters in the neural network increases.
In addition, we observe a linear relation between energy and variance as neural networks scale up (See \spnote{3}).
Accordingly, we adopt an energy-variance extrapolation scheme (Section \ref{sec:method_ext}), which yields our best energy estimate, LAVA with Scaling-law Extrapolation (LAVA-SE).

In all tested cases, we consistently achieve total energy accuracy at sub-kJ/mol level using LAVA-SE together with a robust and reproducible scaling pattern of LAVA (Fig.~\ref{fig:benchmark}b). 
Remarkably, our results align closely with those of the W4 protocol, a CC-based composite procedure extrapolated to the all-electron CCSDTQ5/CBS energy, offering kJ/mol or even ``semi-spectroscopic" accuracy for thermochemistry~\cite{karton_w4-11_2011}. 
With LAVA, applying neural scaling laws alone suffices to surpass the sub-kJ/mol threshold, without the need for \textit{ad hoc} corrections.
Note that this level of accuracy is achieved without relying on any error cancellation, offering a direct and absolute measure of proximity to the exact solution of the Schrödinger equation.

We further assess the efficiency and practicality of our approach from two complementary perspectives, related to other scaling behaviors that emerge as we scale up our calculations.
First we demonstrate that LAVA can maintain favorable convergence speed across both strongly or weakly correlated regimes, taking nitrogen molecule dissociation curve as a representative case in Fig.~\ref{fig:benchmark}c.
Specifically, in order to approach the exact solution (using the LAVA-SE as a reference), the computation scaling with respect to computational runtime remains close to linear with minor fluctuations along the whole curve (Fig.~\ref{fig:benchmark}c lower panel).
As a comparison, couple cluster (CC) theory exhibit significant deterioration in convergence scaling in the strongly correlated regime \cite{chan_state---art_2004} (Fig.~\ref{fig:benchmark}c upper panel).
Moreover, LAVA's scalability can further benefit from parallel computing, whereas CC cannot (see \spnote{2.2}).
Second, we examine LAVA's performance as the molecular size increases. 
These results demonstrate that our approach is able to maintain chemical accuracy with a relatively low computational scaling ($N_e^{5.2}$, where $N_e$ denotes the number of electrons), as shown in Fig.~\ref{fig:benchmark}d.
By comparison, CCSD(T) scales as $N_e^7$, while achieving chemical accuracy may require even higher-order excitations, which impose steeper, if not impractical, scalings.

In the following sections, we address a range of practical chemical challenges using scaled‑up neural‑network wavefunctions.
When a variational guarantee is preferred, we report energies from our largest network; otherwise, we use the extrapolated LAVA‑SE results.

\subsection{Definitive Benchmarks beyond Experimental Limitations}
\begin{figure*}[th]
    \centering
    \includegraphics[width=1.0\textwidth]{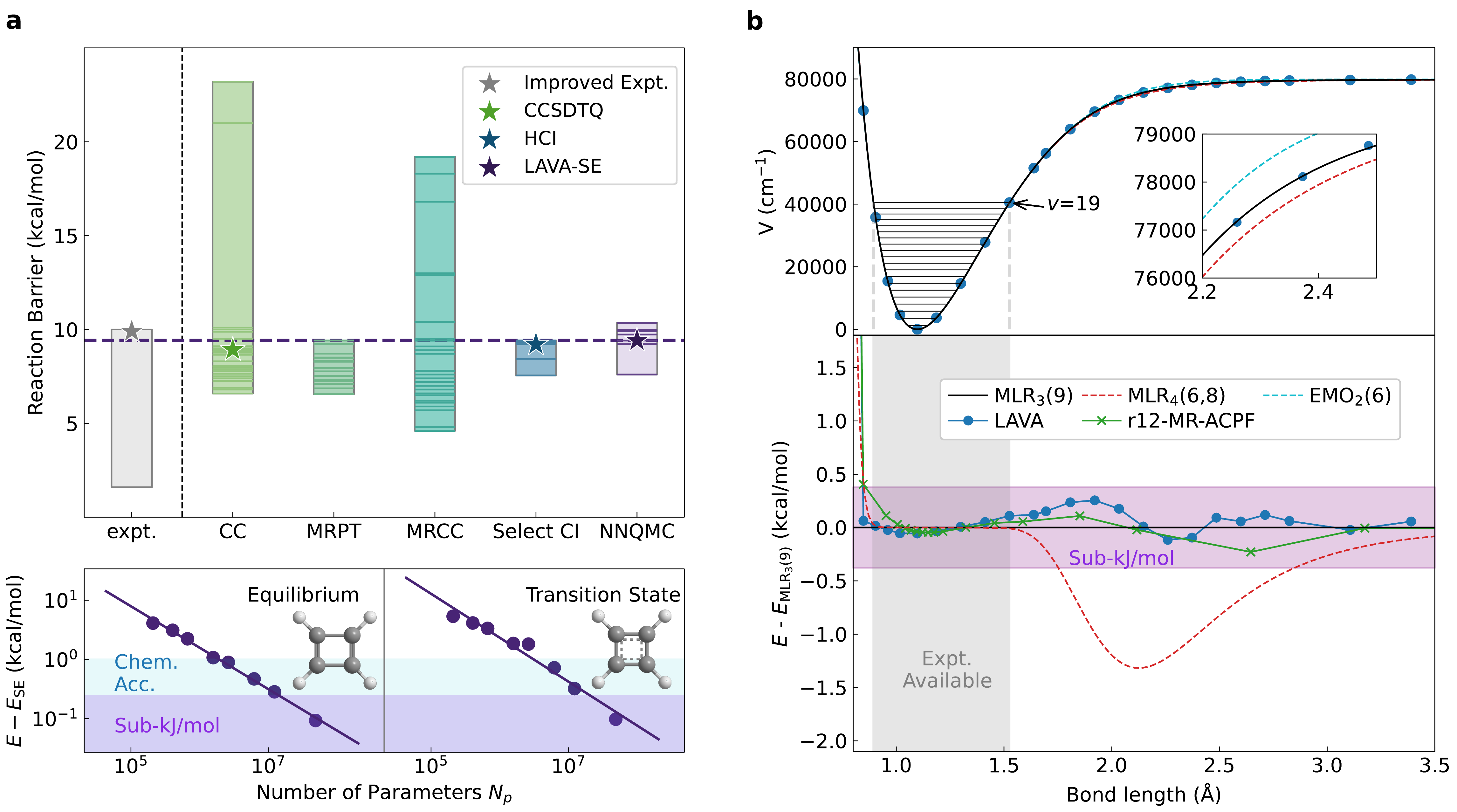}
    \caption{\small {{\bf Definitive LAVA benchmarks complement experiments.}
    \textbf{a.} Upper: Reaction barriers of $\mathrm{C}_4\mathrm{H}_4$ derived from experiment and different theoretical methods.
    Bars represent the spread of reported values, corresponding to each line, within each method family.
    Stars indicate the improved experimental estimate and the best estimates from systematically improvable theories, namely CCSDTQ/aug-cc-pVTZ in CC, HCI/cc-pVTZ extrapolated to the FCI limit in selected CI, and LAVA in NNQMC.
    The dashed line shows the value predicted by LAVA, highlighting the consistency among those stars. 
    Lower: Power-law scaling trends between the number of parameters $N_p$ and energy difference $E-E_{\mathrm{SE}}$, where $E_{\mathrm{SE}}$ are scaling-law extrapolation results.
    \textbf{b.} Upper main panel: The ground-state potential energy curve of N$_2$ dissociation. The analytic potential functions, namely $\text{MLR}_3(9)$, $\text{MLR}_4(6,8)$, and $\text{EMO}_2(6)$, are fitted from vibrational levels $v=0-19$. 
    Upper inset: PECs fitted from different analytic function forms and parameters show a large discrepancy at long bond lengths. 
    Lower main panel: The energy difference between two \textit{ab initio} calculations and three fitted analytic PECs. $\text{MLR}_3(9)$, the surrogate model of LAVA, is the new recommended benchmark.
    The $r_{12}$-MR-ACPF result is shifted together so that its results match high-fidelity experimental data in the equilibrium region (the shaded area). 
    }
    }
    \label{fig:N2_FOOF_C2H4}
\end{figure*}

The scarcity of high-quality experimental data remains a key bottleneck in quantum chemistry, limiting both the development of exchange-correlation (XC) functionals in DFT and the validation of advanced wavefunction methods.
For unstable or non-equilibrium geometries, it is challenging to perform thermochemical experiments to measure their energies and observables.
In such cases, it is particularly valuable to have highly accurate \textit{ab initio} methods to fill the gaps. 
LAVA is well-suited for these challenging regimes, offering a reliable and systematically improvable alternative that serves as a critical complement to experimental data.

As a first demonstration, we study the reaction barrier associated with the automerization of cyclobutadiene ($\mathrm{C}_4\mathrm{H}_4$), a long-standing challenge in quantum chemistry due to its multireference character. 
To date, neither experimentally derived benchmarks nor theoretical predictions can reliably determine the reaction barrier between the rectangular ($D_\text{2h}$) minimum and the square ($D_\text{4h}$) transition state.
The experimental estimate gave a lower bound of 1.6 kcal/mol and an upper bound of 10 kcal/mol, while various theoretical predictions range from 3 to 20 kcal/mol \cite{Monino2022,Lyakh2011,Dang2022, Hatch2025, hermann_deep-neural-network_2020,Ren2023}, as illustrated in Fig.~\ref{fig:N2_FOOF_C2H4}a. 
Here, we establish a definitive benchmark for this transition barrier (9.2 kcal/mol), obtained from LAVA-SE for both $D_\text{2h}$ and $D_\text{4h}$ states.
This result is derived from a sequence of scaled-up neural network wavefunctions and does not rely on error cancellation.
To further validate our result, we establish consensus with improved experimental estimates and the most accurate predictions from different theoretical frameworks, as shown by the star symbols in Fig.~\ref{fig:N2_FOOF_C2H4}a.
For experimental data, we refined the previous rough estimate to a more precise value of 9.9 kcal/mol (See \spnote{6}).
This new estimate agrees well with LAVA and may align even more closely under alternative computational settings during refinement \cite{Whitman1982}.
On the theoretical side, we leverage the systematic improvability of the CC and selected CI families to obtain their best estimates,  which are CCSDTQ/AVTZ \cite{Monino2022} and HCI(20e, 172o)/cc-pVTZ extrapolated to the FCI limit~\cite{Dang2022}, respectively.
Notably, the best estimates from NNQMC, CC, and selected CI---three fundamentally different theoretical approaches---agree with each other within 0.3 kcal/mol, suggesting convergence to the exact reaction barrier.
This consistency highlights LAVA's ability to deliver accurate and reliable solutions for challenging electronic structure problems.

Next, we demonstrate LAVA on the potential energy curve of the nitrogen dimer, a prototypical and long-standing challenge due to its multireference character, and establish a new benchmark, MLR$_3$(9), that surpasses the previous standard MLR$_4$(6,8) in accuracy.
The previous benchmark, labeled as  MLR$_4$(6,8) in  Fig. \ref{fig:N2_FOOF_C2H4}.b, was fitted to experimental vibrational levels $v=0-19$~\cite{le2006accurate}.
These vibrational levels provide a high-accuracy energy benchmark only in the near-equilibrium region, corresponding to bond lengths $r$ between 0.9 to 1.5 $\mathrm{\AA}$ (gray shaded region). 
There are few experimental results for vibrational levels $v$=20-25~\cite{Laher1991}, albeit with a high uncertainty. 
Measuring higher vibrational levels is challenging because the nitrogen dimer becomes unstable in the near-dissociation region. 

Different fitting schemes also give large discrepancies in this region. As shown in the inset of Fig. \ref{fig:N2_FOOF_C2H4}, EMO$_2$(6) and MLR$_4$(6,8) differ by 3.4 kcal/mol in the near-dissociation region, where experimental benchmarks are unavailable. Details of fitting analytic PECS from experimental vibrational levels are given in Supplementary Note 7 and \citet{le2006accurate}.
At the intermediate region, MLR$_4$(6,8)  also shows significant disagreements with multi-reference correlated calculations \cite{Gdanitz1998} and previous NNQMC studies \cite{ferminet,gerard2022goldstandard,gao2024neural, gao2023generalizing, Ren2023}. 

We now provide a definitive benchmark to resolve these discrepancies by performing LAVA calculations across various bond lengths.
Furthermore, we fit an analytic potential curve in the Morse/long-range form \cite{le2006accurate}, namely MLR$_3$(9), based on LAVA data points which also reproduce the experimental vibrational levels $v$=0-19.
Details on this fitting procedure are given in the \spnote{7}. 
This new benchmark MLR$_3$(9) retains the accuracy of the previous benchmark MLR$_4$(6,8) \cite{le2006accurate} for the near-equilibrium region ($0.9\ \mathrm{\AA} <r <1.5\ \mathrm{\AA}$) and the fully dissociation limit($r > 4.0\ \mathrm{\AA}$), while improving the reliability around the near dissociation region. 
In addition, $r_{12}$-MR-ACPF \cite{Gdanitz1998,Flores2005}, an explicitly correlated multireference method with FCI/CBS accuracy, perfectly aligns with our new benchmark MLR$_3$(9) within $1$ kJ/mol difference, after shifting down by about 5 mHa.  However, $r_{12}$-MR-ACPF exhibits exponential scaling with respect to the size of reference space. This computational complexity necessitates careful optimization of both the reference space and basis set, thereby restricting its practical application to small molecular systems.
Together, the reproduction of both the available experimental vibrational levels and the relative energies from $r_{12}$-MR-ACPF confirms the accuracy and reliability of our newly fitted benchmark MLR$_3$(9) curve.

Collectively, the cases in this section demonstrate the reliability and versatility of our approach.
It not only resolves long-standing discrepancies between experimental measurements and theoretical predictions but also provides definitive benchmarks when experiments or conventional theories fall short, thereby establishing a new standard for future benchmark studies.

\subsection{Metastability of Ozone's Ring-Minimum Species}\label{subsec21}
\begin{figure*}[htp]
    \centering
    \includegraphics[width=1.0\textwidth]{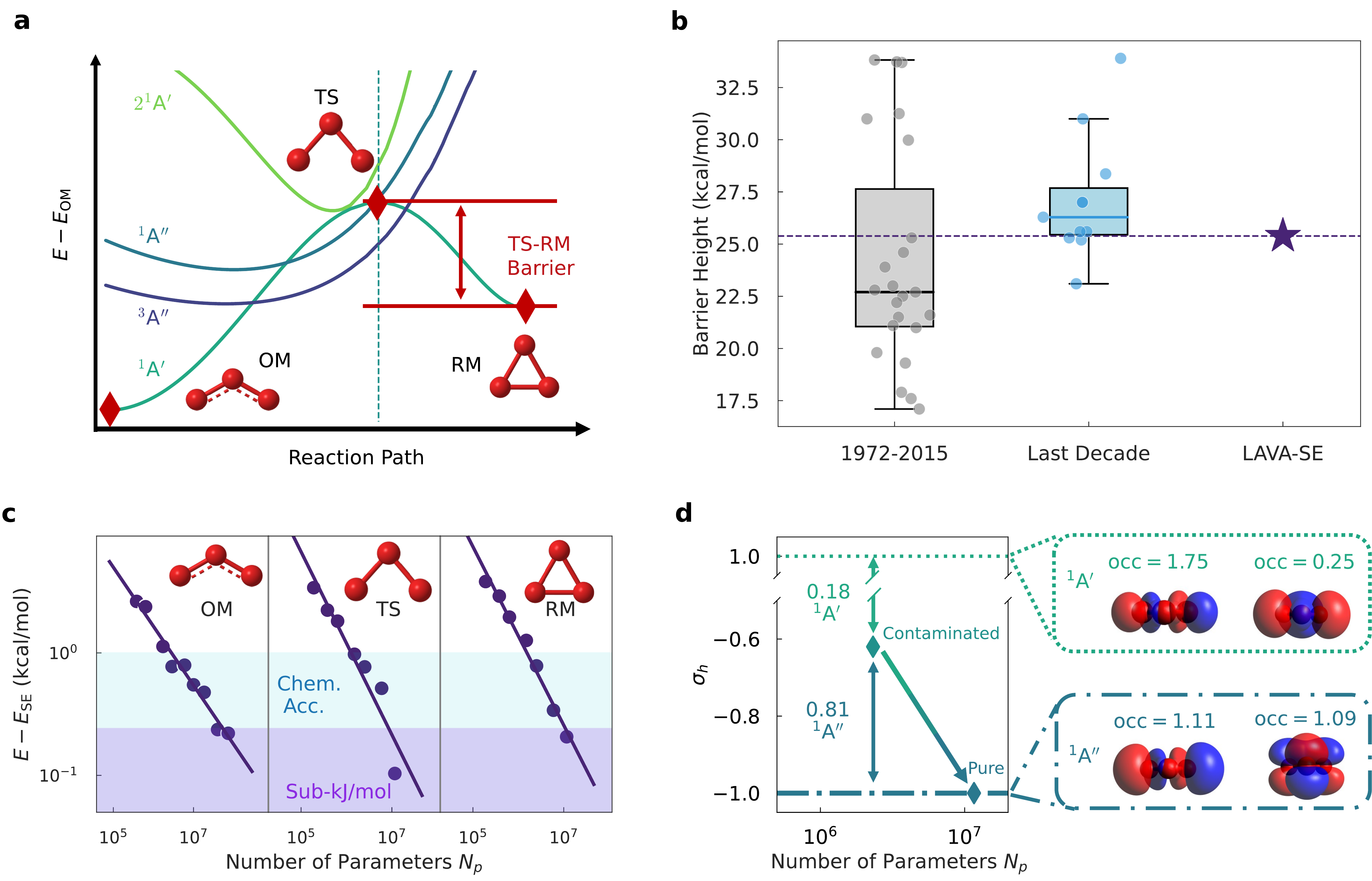}
    \caption{\small {\bf Analysis of ozone bent-cyclic isomerization reaction.} 
    \textbf{a.} A conceptual diagram for $\mathrm{O}_3$ potential energy surfaces of $^1\mathrm{A}^{\prime}$, $^1\mathrm{A}^{\prime\prime}$, $^3\mathrm{A}^{\prime\prime}$, and $2^1\mathrm{A}^{\prime}$ in isosceles triangle geometries, based on XMS-CASPT2 predictions \cite{varga_potential_2017, shu_chempotpy_2023}.
    \textbf{b.} Summary of calculated energy barriers of $\mathrm{O}_3$ isomerization reaction from ring-minimum species to open-ring minimum since 1972.
    Our LAVA-SE estimate, denoted by the purple star, is consistent with most high-accuracy results in the last decade.
    \textbf{c.} Neural scaling laws for OM, TS, RM configurations of ozone, respectively, showing clear power-law decay between energy error and network size.
    The energy error is calculated with respect to LAVA-SE.
    \textbf{d.} Evolution of wavefunction spatial symmetry as the number of parameters increases for the TS configuration. $\sigma_h$ is the expectation of the horizontal reflection operator. 
    With the default network size, LAVA produces a contaminated state as a superposition of $^1$A$^\prime$ state (weight 0.18) and $^1$A$^{\prime\prime}$ state (weight 0.81).
    Enlarged networks yield a pure $^1\mathrm{A}^{\prime\prime}$ state, which is 3 kcal/mol lower in energy compared to the contaminated one.
    For validation purpose, we also generate symmetry-enforced LAVA results for $^1$A$^\prime$ and $^1$A$^{\prime\prime}$, visualizing their natural orbitals with the lowest two occupation numbers that $\mathrm{occ}>0.1$.
    }
    \label{fig:ozone}
\end{figure*}

 Ozone plays a critical role in atmospheric chemistry. 
A long-standing puzzle, however, is the metastability of cyclic ozone, which has been predicted by various theoretical studies, from early \textit{ab initio} work to modern coupled-cluster and multireference methods, since the 1970s~\cite{Wright1973, Shih1974,burton1977theoretical, harding1977b, Xantheas1991, banichevich1993theoretical, atchity1997global, qu2005infrared, chen2011theoretical,theis2016transition, boschen2017correlation, chien2018excited, vitale2020fciqmc, varga2022diabatic}.
Nonetheless, direct experimental evidence remains absent, casting uncertainty on these theoretical predictions.

Here, we tackle this problem by studying the reaction barrier that connects three critical structures, namely the open-ring minimum (OM) (the lowest energy configuration), the equilateral ring minimum (RM, \textit{i.e.}, cyclic ozone), and the transition state (TS) between OM and RM.
Fig.~\ref{fig:ozone}a illustrates the highly complex potential energy surfaces of $\mathrm{O}_3$ spanning those species. 
These surfaces feature numerous intersections, including crossings between states of different spin multiplicities and C$_\text{2v}$ spatial symmetries.
Such complexity presents a significant challenge for electronic structure methods, as accurate modeling requires careful treatment of spin and spatial symmetry constraints and degeneracies.
Despite extensive theoretical efforts over the past half a century, estimates of the barrier height remain far from consensus, differing by more than 15 kcal/mol.
Leveraging neural scaling laws, LAVA predicts a reaction barrier of 24.9 kcal/mol as shown in Fig.~\ref{fig:ozone}c.
Although slightly lower than most values reported from high-accuracy methods in the past decade (Fig.~\ref{fig:ozone}b), this result still supports the kinetic stability of RM, even when accounting for the tunneling effect \cite{chen2011theoretical}.

Additional evidence further supports the accuracy and reliability of our results. 
To begin with, the predictions of LAVA regarding the energy and geometry of OM  are in excellent agreement with experimental results~\cite{Herzberg1966} (See \spnote{8.1}). 
Furthermore, neural scaling laws enable the emergence of physical symmetries without applying explicit constraints to wavefunctions.
As the model size increases, the neural network wavefunction recovers the correct spin multiplicity and spatial symmetry.
Take TS as a showcase,  LAVA guides the neural network wavefunctions toward the correct $^3\mathrm{A}^{\prime\prime}$ ground state without any constraints (see \spnote{8.6}). Additionally, when we enforce singlet spin symmetry using a penalty-based method \cite{li_spin-symmetry-enforced_2024}, LAVA identifies the $^1\mathrm{A}^{\prime\prime}$ ground state, as evidenced by the improved spatial symmetry depicted in Fig.~\ref{fig:ozone}d.

Overall, such a capability of scaling up towards the correct ground state is essential for reliable ground-state characterization in strongly correlated regimes, where multiple near-degenerate states often engage in competition.

\subsection{Beyond Energies}
\begin{figure}[htp]
    \centering
    \includegraphics[width=\linewidth]{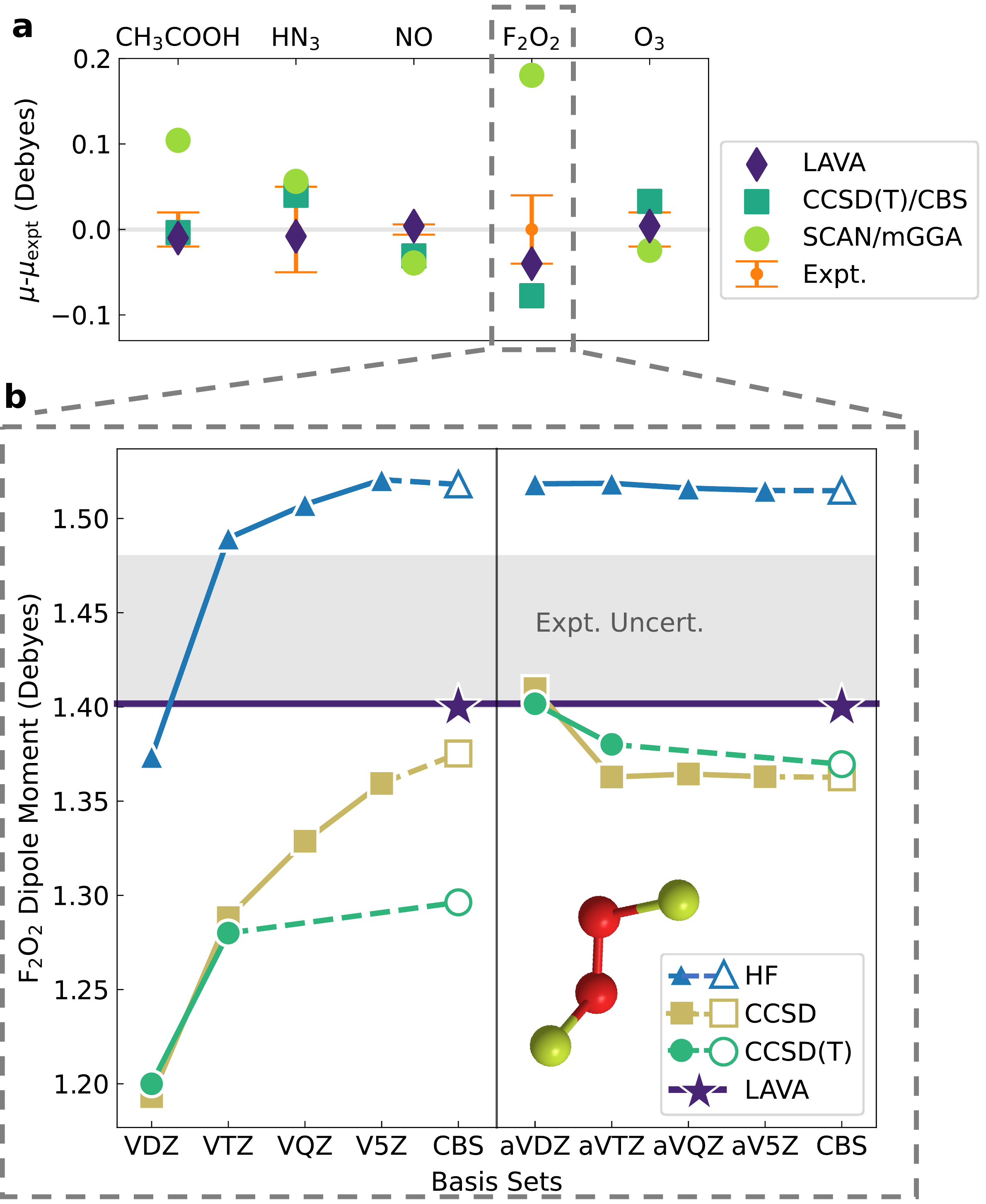}
    \caption{\small {\bf LAVA delivers reliable dipole moments.}
    \textbf{a.} Dipole moments $\mu$ derived from our wavefunctions match experimental values $\mu_{\mathrm{expt}}$ \cite{NIST_CCCBDB}.
    For comparison, CCSD(T)/CBS results from \citet{hait_how_2018} are also visualized, which fall out of the experimental range for NO, O$_3$, and F$_2$O$_2$. For all molecules, the SCAN density functional results \cite{hait_how_2018} fail to match experimental references.
    \textbf{b.} Basis set convergence of dipole moment of F$_2$O$_2$ for HF and coupled-cluster methods, together with experimental benchmark and LAVA estimates for comparison. Dipole moments in Debyes are plotted against correlated-consistent basis sets (cc-pVXZ) and augmented counterparts (aug-cc-pVXZ) for $\mathrm{X}=\mathrm{D},\mathrm{T},\mathrm{Q},5$.
    CBS extrapolation scheme follows \spnote{6.2}.
    The missing points indicate calculations infeasible with available computational resources.
    Results of coupled-cluster methods exhibit strong basis set dependence and severe memory bottlenecks, while LAVA achieves experimental-level accuracy.
    }
    \label{fig:dipole}
\end{figure}
Beyond accurate absolute energies, LAVA also produces near-exact wavefunctions in the CBS limit, benefiting from its first-quantized nature.
Consequently, it delivers FCI/CBS-quality physical observables including electron densities and dipole moments.
Conventional quantum chemistry methods, by contrast, remain limited by the finite basis set approximation, despite decades of effort toward developing systematically improvable basis sets \cite{Dunning1989}.
While relative energies tend to converge rapidly with basis set size \cite{Feller2011,Karton2020}, the convergence behavior of electron densities and density-derived properties toward the CBS limit remains less well understood \cite{hait_how_2018, Hait2021}.

We assess LAVA’s performance on dipole moments, a critical property reflecting molecular charge distribution and polarity.
Specifically, we compare LAVA predictions with highly accurate references from spectroscopy experiments \cite{NIST_CCCBDB} (Fig.~\ref{fig:dipole}a).
Across all the molecules examined, LAVA shows excellent agreement with experimental data, falling within the bounds of experimental uncertainty.
In contrast, CCSD(T)/CBS results deviate from these bounds for molecules with significant multireference character ({see \sptable{17}), namely $\text{O}_3$ and $\text{F}_2\text{O}_2$.
One could, in principle, move beyond CCSD(T)/CBS to CCSDT(Q)/CBS for better accuracy, but such calculations are computationally intractable due to the steep scaling of coupled cluster methods.
Moreover, dipole moments are notoriously sensitive to the choice of basis set and converge slowly toward the CBS limit.
Fig.~\ref{fig:dipole}b exemplifies this behavior for $\text{F}_2\text{O}_2$, where neither CCSD nor CCSD(T) reaches experimental accuracy when extrapolated along either the correlated-consistent basis set sequence (cc-pVXZ) or its augmented counterpart (aug-cc-pVXZ). Memory demand grows rapidly with increasing basis set size, emerging as the computational bottleneck: CPU-based PySCF~\cite{Sun2020} is limited to the aVTZ basis set, while GPU-accelerated ByteQC \cite{Guo2025} can accommodate the larger aV5Z basis set.
Notably, LAVA also produces more accurate energy than CC-based composite theory W4 for molecules with strong multireference character, as discussed \spnote{4.3}.

Moreover, with scaled-up neural networks, LAVA is also able to produce high-quality electron densities \cite{cheng2025highly}.
Together with benchmark-quality energies, this provides critical reference data for developing better density-functional approximations that seek to recover the exact energy from the exact density \cite{Medvedev2017}.

\section{Discussion}\label{discussion}
This study demonstrates how LAVA-powered neural scaling laws can systematically and practically approach exact solutions to many-electron Schrödinger equations.
We generate high-quality benchmarks for various molecules, overcoming the issues of uncertainties and inaccuracies in experimental data.
Additionally, we confirm the metastability of ring-minimum ozone from a theoretical perspective.

We are now able to provide FCI/CBS quality wavefunctions with an attractive computational scaling.
This enables the creation of a reliable benchmark dataset of both energies and densities, complementing scarce experimental references.
Such datasets can facilitate data-driven development of next-generation XC functionals in DFT and rigorous validation of other state-of-the-art wavefunction methods, opening new chapters in high-accuracy quantum chemistry and its applications.  
Besides its favorable computational scaling and embarrassingly parallelism, LAVA also benefits from other recent algorithmic progresses in the field of NNQMC \cite{scherbela2025accurate, peng2025analysis, fu2025local}, which further reduce the computational cost significantly.
Moreover, with rapid advances in AI hardware and optimization techniques, we anticipate that this approach will become increasingly practical and scalable, enabling broader applications to larger systems in quantum chemistry and beyond.

Overall, the synergy of AI and quantum chemistry offers a powerful route to solving complex electronic structure problems with near-exact solutions, unlocking transformative applications in catalysis, materials science, and drug discovery. 
While challenges remain, neural scaling laws establish a robust foundation for advancing accuracy and scalability across increasingly complex systems in \textit{ab initio} quantum chemistry.

\section{Methods}\label{sec3}

\subsection{Variational Optimization Framework}\label{sec:method_vmc}

In this work, we focus on the time-independent electronic Schr\"{o}dinger equation within the Born-Oppenheimer approximation:
\begin{align}
    \hat{H} \psi&\left({x}_1, \ldots, {x}_N\right)=E \psi\left({x}_1, \ldots, {x}_N\right),\\
    \hat{H}= & -\frac{1}{2} \sum_i \Delta_i+\sum_{i>j} \frac{1}{\norm{\mathbf{r}_i-\mathbf{r}_j}}\nonumber \\ 
        & -\sum_{i I} \frac{Z_I}{\left|\mathbf{r}_i-\mathbf{R}_I\right|}+\sum_{I>J} \frac{Z_I Z_J}{\left|\mathbf{R}_I-\mathbf{R}_J\right|},
\end{align}
where for the $i$-th electron ($i\in\{1,2,\cdots,N\}$), ${x}_i=\{\mathbf{r}_i,\sigma_i\}$ consists of the coordinate of electron $\mathbf{r}_i\in\mathbb{R}^3$ and its spin $\sigma_i\in\{1,-1\}$, and for the $I$-th nucleus ($I\in\{1,2,\cdots,M\}$), we denote the charge as $Z_I$ and its fixed position as $\mathbf{R}_I$. Obeying the spin-statistics theorem, a many-electron wavefunction is antisymmetric under an exchange of electrons.

As an \textit{ab initio} method, variational Monte Carlo (VMC) directly solves the following optimization problem to approximate the ground state of a many-electron quantum system:
\begin{equation}
\begin{aligned}
\mathop{\mathrm{min}}\limits_{\theta}&\qquad\frac{\langle\psi_\theta|\hat{H}|\psi_\theta\rangle}{\langle\psi_\theta|\psi_\theta\rangle},\\
\mathrm{s.t.}&\qquad\psi(\cdots,{x}_i,\cdots,{x}_j,\cdots)\\\ &\qquad=-\psi(\cdots,{x}_j,\cdots,{x}_i,\cdots),\\
\ &\qquad\forall i,j\in\{1,2,\cdots,N\},
\end{aligned}
\end{equation}
and $\theta$ are ansatz parameters for representing neural network wavefunction $\psi_\theta$. VMC uses Monte Carlo methods to evaluate the expected value of the total energy 
\begin{equation}
\begin{aligned}
E_{\mathrm {tot}} 
& =\frac{\langle\psi_\theta|\hat{H}|\psi_\theta\rangle}{\langle\psi_\theta|\psi_\theta\rangle} = \int\frac{\hat{H}\psi_\theta(x)}{\psi_\theta(x)}\frac{\psi_\theta^2(x)}{\langle\psi_\theta|\psi_\theta\rangle} \dd x \\
& = \mathbb{E}_{x\sim p}[E_L],
\end{aligned}
\end{equation}
where $p(x)=\frac{\psi^2_\theta(x)}{\langle\psi_\theta|\psi_\theta\rangle}$ and $E_L=\frac{\hat{H}\psi_\theta(x)}{\psi_\theta(x)}$ is so-called local energy. Variational optimization is then performed to find the best approximation of the ground state wavefunction in the ansatz space.

From another perspective, solving the ground state of the time-independent Schr\"odinger equation can be seen as finding the lowest eigenstate of the Hermitian operator $\hat{H}$, which can be achieved by the power method \cite{golub_matrix_2013}. By repeatedly applying a propagator 
to an arbitrary trial state $\psi$, one can eventually get the ground state $\psi_0$, as long as $\psi$ and $\psi_0$ are not orthogonal. 
The propagator can be any operator that decays all the other eigenstates of the Hamiltonian $\hat{H}$ while retaining the one with the lowest eigenvalue. 
When we use the exponential form $e^{-\tau (\hat{H} - E_T)}$ as propagator, the power method is equivalent to imaginary time evolution under the Wick-rotated Schr\"odinger equation \cite{zee_quantum_2010}. Here, $E_T<0$ is a scalar close to the ground state energy for normalization. 

We consider the linear propagator
\begin{equation}
    \hat{U}(\tau) = 1 - \tau (\hat{H}-E_T).
\end{equation}
For the finite time inteval $\tau$, as $n\tau\rightarrow +\infty$, where integer $n$ is the number of time steps, we have $\psi_0 \propto \hat{U}^n(\tau)\psi$.

From the perspective of the power method, VMC can be deemed as an alternating application of two operators: the propagator to evolve towards the ground state, and a projection operator that maps the propagated state back to the ansatz space. Given an ansatz space parametrized by $\theta\in\Omega$, where $\Omega$ is the parameter space, we define the projection operator $\hat{P}_{\mathcal{D}_1, \mathcal{D}_2}({\theta,\eta})$ that projects an arbitrary wavefunction $\phi$ to the ansatz space in the vicinity of $\theta$, with radius $\eta$, while keeping the projected wavefunction normalized, under parameter space metric $\mathcal{D}_1$ and Hilbert space metric $\mathcal{D}_2$:
\begin{equation}
\hat{P}_{\mathcal{D}_1, \mathcal{D}_2}({\theta,\eta})\, \phi = \sqrt{\frac{{\langle\psi_{\theta}|\psi_{\theta}\rangle}}{{\langle\psi_{\theta^*}|\psi_{\theta^*}\rangle}}}\psi_{\theta^*},
\end{equation}
where
\begin{equation}
\theta^* = f_{\mathcal{D}_1,\mathcal{D}_2}(\theta,\eta)\phi \triangleq \argmin_{\substack {\theta^\prime\in\Omega\\\mathcal{D}_1(\theta, \theta') \leq \eta}} \mathcal{D}_2(\phi, \psi_{\theta'}).
\end{equation}

In the case of VMC with natural gradient descent (or stochastic reconfiguration), $\mathcal{D}_1$ is the metric induced by the quantum geometric tensor $\mathcal{F}$, and $\mathcal{D}_2$ is the Kullback-Leibler (KL) divergence.

At step $n$, the iteration of VMC, starting from parameters $\theta_{n}$, can be written as
\begin{equation}
\begin{aligned}
&{\theta_{n+1}}={\theta_n}+\eta_n g,\\
&g=\lim_{\tau\rightarrow 0+}
\frac{f_{\mathcal{F}, \mathrm{KL}}({\theta_n,\eta_n\tau})\,
\hat{U}(\tau)\,
\psi_{\theta_n}-{\theta_n}}{\eta_n\tau},
\end{aligned}
\end{equation}
where $\eta_n$ is the learning rate, which can vary with the iteration.

Neural network-based variational Monte Carlo \cite{ferminet,glehn2023a,schatzle2023deepqmc} utilizes a neural network-based ansatz to parameterize the many-body wavefunction, enabling the accurate capture of complex electronic correlations that are often challenging for traditional quantum chemistry methods. Following \citet{li_computational_2024}, we use the LapNet ansatz and the Forward Laplacian computational framework. 

Instead of directly minimizing total energy, we develop the Lookahead Variational Algorithm (LAVA) for optimization, which is described in more detail below.

\subsection{\name: the Lookahead Variational Algorithm} \label{sec:method_lava}

LAVA is designed to optimize parameterized wavefunctions by combining the principles of imaginary time evolution (ITE) and variational optimization. We describe LAVA's implementation in Algorithm \ref{alg:LAVA}. Inspired by the Lookahead algorithm \cite{10.5555/3454287.3455148}, our approach calculates the Lookahead direction in Hilbert space following the discretized ITE trajectory (Algorithm \ref{alg:LAVA} line 6-12) and updates the variational ansatz through a projection mechanism ensuring that optimization remains confined to the variational manifold (Algorithm \ref{alg:LAVA} line 13-14). 

From this perspective, LAVA ``looks ahead" at the space of iteratively propagated states. Instead of directly using the projected propagation for the next step, in iteration $n$, LAVA first constructs temporary state $\psi_{\mathrm{temp}}$ as
\begin{equation}
\psi_{\mathrm{temp}} = 
\hat{P}_{\mathcal{F}, \mathrm{KL}}({\theta_n,\eta_\mathrm{temp}})\,
\hat{U}(\tau_\mathrm{temp})\,
\psi_{\theta_n},
\end{equation}
where $\theta_n$ denotes the current parameters, and $\eta_{\mathrm{temp}}$, $\tau_{\mathrm{temp}}$ are the projection radius and the small time interval used in this intermediate step, respectively.

From this intermediate $\psi_{\mathrm{temp}}$, LAVA moves along a Lookahead direction:
\begin{equation}
    \Delta\psi=-\tau\hat{H}(\psi_{\mathrm{temp}}+\psi_{\theta_n})+2\tau E_T\psi_{\theta_n}.
\end{equation}
Since
\begin{equation}
\Delta\psi\propto\frac{1}{2E_T}\hat{H}(\psi_{\mathrm{temp}}+\psi_{\theta_n})-\psi_{\theta_n},
\end{equation}
to get $\psi_{\theta_{n+1}}$, LAVA applies a projection operator $\hat{P}_{\mathcal{F}, \mathrm{SM_1}}({\theta_n,\eta})$ on 
\begin{equation}
\psi^\prime=\frac{1}{E_T}\hat{H}(\psi_{\theta_\mathrm{temp}}+\psi_{\theta_n}).
\end{equation}
For this projection $\hat{P}_{\mathcal{F}, \mathrm{SM_1}}({\theta_n,\eta})$, we use $L_1$ score matching (SM):
\begin{equation}
\mathrm{SM}_1(\phi,\psi;p)=\mathbb{E}_{x\sim p}[\|\nabla_x\ln|\phi(x)|-\nabla_x\ln|\psi(x)|\|_1].
\end{equation}
Then,
\begin{equation}
\label{eq:LAVA_concept}
\psi_{\theta_{n+1}} = \hat{P}_{\mathcal{F}, \mathrm{SM_1}}({\theta_n,\eta})\,
\psi^\prime.
\end{equation}

For the actual implementation, our projected propagation uses natural gradient descent with Kronecker-Factored Approximate Curvature (known as KFAC) \cite{10.5555/3045118.3045374}, in which a block-diagonal $\mathcal{F}_{\mathrm{KFAC}}$ approximates the
quantum geometric tensor $\mathcal{F}$.
At iteration $n$, LAVA first performs a VMC step to get intermediate parameters $\theta_{\mathrm{temp}}$:
\begin{equation}
\begin{aligned}
&    \theta_{\mathrm{temp}}=\theta_{n}-\eta_\mathrm{temp} \mathcal{F}^{-1}_{\mathrm{KFAC}} 
    g,\\
&g=\mathbb{E}_{x\sim\psi_{\theta_n}^2} [(E_L(x) - E_{\mathrm{tot}}) \nabla_{\theta_{n}} \mathrm{ln} |\psi_{\theta_{n}}(x)|],
\end{aligned}
\end{equation}
where $E_\mathrm{tot}$ is the average total energy and $\eta_{\mathrm{temp}}$ is the learning rate for this intermediate step. An unnormalized temporary state $\psi^\prime$ is calculated by 
\begin{equation}
\psi^\prime = -\hat{H}\left(\psi_{\theta_\mathrm{temp}}\sqrt{\frac{\langle\psi_{\theta_n}|\psi_{\theta_n}\rangle}{\langle \psi_{\theta_\mathrm{temp}}| \psi_{\theta_\mathrm{temp}}\rangle}} + \psi_{\theta_n}\right).
\end{equation}
The gradient direction of LAVA then follows
\begin{equation}
\begin{aligned}
g_n&=\nabla_{\theta_n}\mathrm{SM}_1(\psi^\prime,\psi_{\theta_n};\psi_{\theta_{\mathrm{temp}}}^2)\\
    &= \mathbb{E}_{x\sim \psi_{\theta_{\mathrm{temp}}}^2}[\left\langle f(x), 
    \nabla_{\theta_{n}}\nabla_{r}\mathrm{ln}|\psi_{\theta_{n}}(x)|\right\rangle],
\end{aligned}
\end{equation}
\begin{equation}
f(x)=-\mathrm{sgn}\left(\nabla_{r} \ln \left|\frac{\psi^\prime(x)}{\psi_{\theta_{n}}(x)}\right|\right).
\end{equation}
LAVA feeds the gradient into the Adam-KFAC optimizer to update the parameters. Unlike \citet{Izadi2020OptimizationOG}, we directly apply Adam \cite{Kingma2014AdamAM} to KFAC preconditioned gradient:
\begin{equation}
\theta_{n+1}=\theta_n-\mathrm{Adam}(\mathcal{F}_\mathrm{KFAC}^{-1}g_n).
\end{equation}

\begin{algorithm*}
\caption{LAVA}\label{alg:LAVA}
\begin{algorithmic}[1]
\Require initial parameters $\theta$, samples $\{x^{(i)}\}_{i=1}^B$
\State $n \gets 0$
\State $m_0\gets 0$
\State $v_0\gets 0$
\While{$n\leq N$}
\State update $x^{(i)}$ by sampling from $\frac{\psi^2_{\theta}}{\langle\psi_{\theta}|\psi_{\theta}\rangle}$
\State $E_L(x^{(i)}) \gets \frac{\hat{H}\psi_{\theta}(x^{(i)})}{\psi_{\theta}(x^{(i)})}$
\State $E_{\mathrm {tot}} \gets \frac{1}{B}\sum\limits_{i=1}^B E_L(x^{(i)})$
\State $g\gets \frac{1}{B}\sum\limits_{i=1}^B(E_L(x^{(i)})-E_{\mathrm{tot}})\nabla_{\theta}\mathrm{ln}|\psi_{\theta}(x^{(i)})|$
\State $\theta_{\mathrm{temp}}\gets \theta-\eta_{\mathrm{temp}}\mathcal{F}^{-1}_\mathrm{KFAC}g$
\State update $x^{(i)}$ by sampling from $\frac{\psi^2_{\theta_{\mathrm{temp}}}}{\langle\psi_{\theta_{\mathrm{temp}}}|\psi_{\theta_{\mathrm{temp}}}\rangle}$
\State $C \gets\sqrt{\frac{\langle \psi_{\theta}|\psi_{\theta}\rangle}{\langle\psi_{\theta_{\mathrm{temp}}}|\psi_{\theta_{\mathrm{temp}}}\rangle}}$
\State $E_L^\prime(x^{(i)})\gets \frac{\hat{H}\left(\psi_{\theta}(x^{(i)})+C\psi_{\theta_{\mathrm{temp}}}(x^{(i)})\right)}{\psi_{\theta}(x^{(i)})}$

\State $\begin{aligned}
        g\gets
        -\frac{1}{B}\sum\limits_{i=1}^B\left\langle \mathrm{sgn}(E_L^\prime(x^{(i)}))\mathrm{sgn}(\nabla_{r^{(i)}}E_L^\prime(x^{(i)}))\right.,
        \left.\nabla_{\theta}\nabla_{r^{(i)}}\mathrm{ln}|\psi_{\theta}(x^{(i)})|\right\rangle
        \end{aligned}$

\State $\theta \gets \theta- \mathrm{Adam}(\mathcal{F}^{-1}_{\mathrm{KFAC}}g)$
\State $n\gets n+1$
\EndWhile\\
\Return $\theta,\ \{x^{(i)}\}_{i=1}^B$
\end{algorithmic}
\end{algorithm*}

\subsection{Neural Scaling Laws} \label{sec:method_scale}

The central premise of our scaling laws establishes that, with LAVA, the energy error of trained models decays following a power law trend with respect to the number of neural network parameters $N_p$ governing expressivity:
\begin{equation}
    E-E_0=\alpha N_p^{-\beta},
\end{equation}
where $\alpha$ and $\beta$ are system-dependent variables and $E_0$ is the exact ground state energy. We monotonically increase network width and the number of determinants during this scaling-up process. For $11$ systems with reliable experimental benchmarks, the average $r^2$ of ordinary least squares (OLS) regression on a logarithmic scale is larger than $0.95$, and residual diagnostics (White test, $p>0.2$) reveals no significant heteroscedasticity. For the details, see \spnote{2}.

We also observe the scaling laws of local energy variance $V=\langle(\hat{H}-E)^2\rangle$:
\begin{equation}
    V=\alpha_v N_p^{-\beta_v},
\end{equation}
where $\alpha_v$ and $\beta_v$ are system-dependent variables. Under the zero-variance principle of quantum Monte Carlo, our estimated energy approximates the exact ground state energy $E_{0}$ as $V\rightarrow 0$. Since variance estimation requires no exact reference data, unlike error measurement, verifying the relationship is possible for any system where LAVA calculations are available. We confirmed this relationship across various molecular systems by linear regression on a logarithmic scale, and the average $r^2$ is larger than $0.97$. Still, OLS is used since the White test shows $p>0.2$. The detailed results are in \spnote{2}, and model architectures and training scheme are listed in \spnote{1.1}.

\subsection{Extrapolation Scheme}\label{sec:method_ext}
For practically trained models, we empirically observed the following relationship between local energy variance and energy: 
\begin{equation}
E= kV+b,
\end{equation}
where $k$ and $b$ are system-dependent variables. This relationship is similar to variance-energy extrapolation in \citet{Fu_2024}, which utilizes training data of a fixed model instead of optimized networks of different capacities.

Since theoretically $\mathop{\mathrm{lim}}\limits_{V\rightarrow 0}E=E_{0}$ and empirically $\mathop{\mathrm{lim}}\limits_{N_p\rightarrow+\infty}V=0$, we follow \citet{Fu_2024} and use $b$ as our asymptotic estimates.
For various molecular systems, residual diagnostics (White test, $p>0.2$) support the use of ordinary least squares regression, and the average OLS $r^2$ for linear regression lines is greater than $0.99$. For systems with experimental reference values, the errors in our extrapolation fall within the experimental uncertainty. Combining neural scaling laws and extrapolation enables the estimation of threshold computational resources to reach sub-kJ/mol accuracy.

\backmatter
\bmhead{Data Availability}

All data supporting the findings of this study are available within the Supplementary Information.

\bmhead{Acknowledgements}

We thank Chenyang Li, Xuefei Xu, Yinan Shu, Donald G. Truhlar for the insight discussion.
We thank Zigeng Huang, Qiming Sun, Xiaojie Wu, and the rest of the ByteDance Seed Group for their inspiring ideas and encouragement. 
We also thank Hang Li for his guidance and support.
L.W. is supported by National Science and Technology Major Project (2022ZD0114902) and National Science Foundation of China (NSFC92470123, NSFC62276005). D.H. is supported by National Science Foundation of China (NSFC62376007).
W.A.G. thanks the US NSF (CBET 2311117) for support.
J.C. is supported by the National Key R\&D Program of China (2021YFA1400500) and National Science Foundation of China (12334003).

\bmhead{Competing interests}
The authors declare no competing interests.

\bibliography{sn-bibliography}

\end{document}

\title{Supplementary Information}

\maketitle
\tableofcontents
\newpage

\section{Look-Ahead Variational Algorithm (LAVA)}
Now we illustrate in detail the effectiveness of our LAVA-based approach.
First, as an optimization framework, LAVA achieves much better accuracy, especially for systems challenging for QMC calculations.
For instance, for systems involving heavier atoms, such as Sulfor and Chlorine, NNQMC approaches struggle due to the large number of inner electrons.
Nonetheless, LAVA can achieve significant accuracy improvement and successfully reach chemical accuracy.

The same holds for strongly correlated systems such as $\text{O}_3$. We directly compare LAVA and NNVMC in \Cref{sec:vmc}.

\subsection{Schema and hyperparameters}

All reported algorithms were implemented in JAX \cite{jax2018github_3_13}. The LAVA distributed calculations used A800 GPUs. The hyperparameters are in \Cref{table:default params}. In particular, for benzene, we use $2$ MCMC blocks to ensure good mixing. For large networks, we observed that for certain systems, $3\text{e}5$ iterations are not sufficient for good convergence, so we increased it to $4\text{e}5$ or $5\text{e}5$ if local energy variance at least decreased by half during the last $5\text{e}4$ iterations. As for denoting network configurations, we use 4-tuples that consist of depth, width, the number of attention heads, and the number of determinants in the following sections.

\begin{table*}[htbp]
    \centering
    \caption{Default hyperparameters.}
    \begin{tabular}{ccc}
    \toprule\midrule
        & Parameter & Value \\
    \midrule
    \multirow{6}{*}{Training}& 
    Optimizer & Adam-KFAC \\
    & Optimizer for intermediate steps & KFAC \\
    & Iterations & 3e5 \\
    & Batch size & 4096 \\
    & $\eta$ at iteration $t$ & $5\text{e-4}/(1+\frac{t}{t_{\mathrm{delay}}})$ \\
    & $\eta_{\mathrm{temp}}$ at iteration $t$ & $5\text{e-3}\cdot\mathrm{min}\{1,\frac{t}{t_{\mathrm{warmup}}}\}/(1+\frac{\mathrm{max}\{t,t_{\mathrm{warmup}}\}}{t_{\mathrm{delay}}})$ \\
    & Learning rate decay $t_{\mathrm{delay}}$ & 1e4\\
    & Linear warmup iterations $t_\mathrm{warmup}$ & 1e5 \\
    & Local energy clipping & 5.0 \\
    \midrule
    \multirow{2}{*}{Inference} & Iterations for energy evaluation & $3\text{e}4$ \\
    &Iterations for dipole evaluation & $1\text{e}6$ \\ 
    \midrule
    \multirow{2}{*}{Pretraining} & Optimizer & LAMB \\
    & Iterations & 2e4 \\
    & Basis set & aug-cc-pVDZ \\
    & Learning rate &3e-4\\
    \midrule
    \multirow{2}{*}{MCMC} & Decorrelation steps & 30 \\
    & Proposal standard deviation & 0.02 \\
    & Blocks & 1 \\
    \midrule
    \multirow{4}{*}{KFAC} & Norm constraint & $1\text{e-7}\cdot\left(\mathrm{min}\{1,\frac{t}{t_{\mathrm{warmup}}}\}/(1+\frac{\mathrm{max}\{t,t_{\mathrm{warmup}}\}}{t_{\mathrm{delay}}})\right)^2$ \\
    & Damping & 1e-3 \\
    & Momentum & 0 \\
    & Covariance moving average decay & 0.95 \\
    \midrule
    \multirow{4}{*}{Adam-KFAC} & Norm constraint & 1e-6\ \\
    & Damping & 5e-3 \\
    & Momentum decay rate $\beta_1$ & 0.9 \\
    & Squared gradients decay rate $\beta_2$ & 0.99 \\
    & $\epsilon$ & 1e-10 \\
    & Covariance moving average decay & 0.95 \\
    \midrule\bottomrule
    \end{tabular}
    \vspace{-0.3cm}
    \label{table:default params}
\end{table*}

\subsection{Energy comparison against VMC}
\label{sec:vmc}
For the LAVA ablation study, we compared energy results with default settings for several systems. LapNet baseline for NNVMC optimization uses default hyperparameters as described in \cite{li_computational_2024}, except for the number of iterations, which is increased to $1\text{e}6$ (specifically $1.35\text{e}6$ for benzene) so as to make the total computational time costs similar. In the following sections, we refer to the LapNet baseline as NNVMC. For both methods, we use the default network, denoted as $(4,256,4,32)$, with $256$ hidden dimensions, $4$ attention heads, $4$ layers, and $32$ determinants (in particular $64$ determinants for ozone open-ring minimum species). Results are shown in \Cref{table:ablation_vmc}. LAVA performs significantly better in absolute energy compared to the NNVMC baseline with comparable time costs.

\begin{table}[htbp]

    \centering
    \caption{LAVA vs. NNVMC energy results with similar time costs.}
    \begin{threeparttable}
    \begin{tabular}{cS[table-format=5.7]S[table-format=5.7]}
    \toprule\midrule
     Molecule  & \text{LAVA (Ha)}& \text{NNVMC}\tnote{a}~\text{ (Ha)}\\
    \midrule
    Cyclobutadiene (Rectangle) & -154.68606(3)	&-154.68413(4)\\
    Cyclobutadiene (Square) &-154.66953(3)	&-154.66797(5)\\
    $\text{O}_3$ &-225.43557(3)&	-225.43433(5)\\
    $\text{SO}_2$ &-548.65702(8)	&-548.6484(1)\\
    $\text{Cl}_2$& -920.3889(2)	&-920.3832(2)\\
    $\text{F}_2\text{O}_2$&-349.84415(6)	&-349.8336(1)\\
    Benzene & -232.24689(5)	&-232.24388(6)\\
    \midrule\bottomrule
    \end{tabular}
    \label{table:ablation_vmc}
    \begin{tablenotes}
    \item[a] LapNet is used as the baseline.
    \end{tablenotes}
    \end{threeparttable}
\end{table}

\section{Neural scaling laws}
The second important component of our approach is the neural scaling laws towards exactness, which is universal across all the tested systems. 
As investigated in detail in artificial intelligence literature, neural scaling laws state that with larger model size and compute resources, the performance of deep neural networks improves predictably following a power law.
In the main text, we have demonstrated such a phenomenon among a number of systems with reliable experimental benchmarks. And the scaling law of our LAVA-based approach is significantly more efficient than the couple-cluster method in terms of approaching exactness.

Moreover, across all the neural scaling law patterns, we discovered a simple but evident linear relationship between the energy expectation and the local energy variance of different neural network wavefunctions, which can be used to devise a straightforward extrapolation scheme. We discuss the scheme in \Cref{sec:ext}. The extrapolated energies are denoted as $E_{\mathrm{SE}}$.

\begin{figure}[htp]
    \centering 
    \caption{Power-law scaling trends between energy difference $E-E_{\mathrm{SE}}$ and the number of parameters $N_p$.}
    \includegraphics[width=1.0\textwidth]{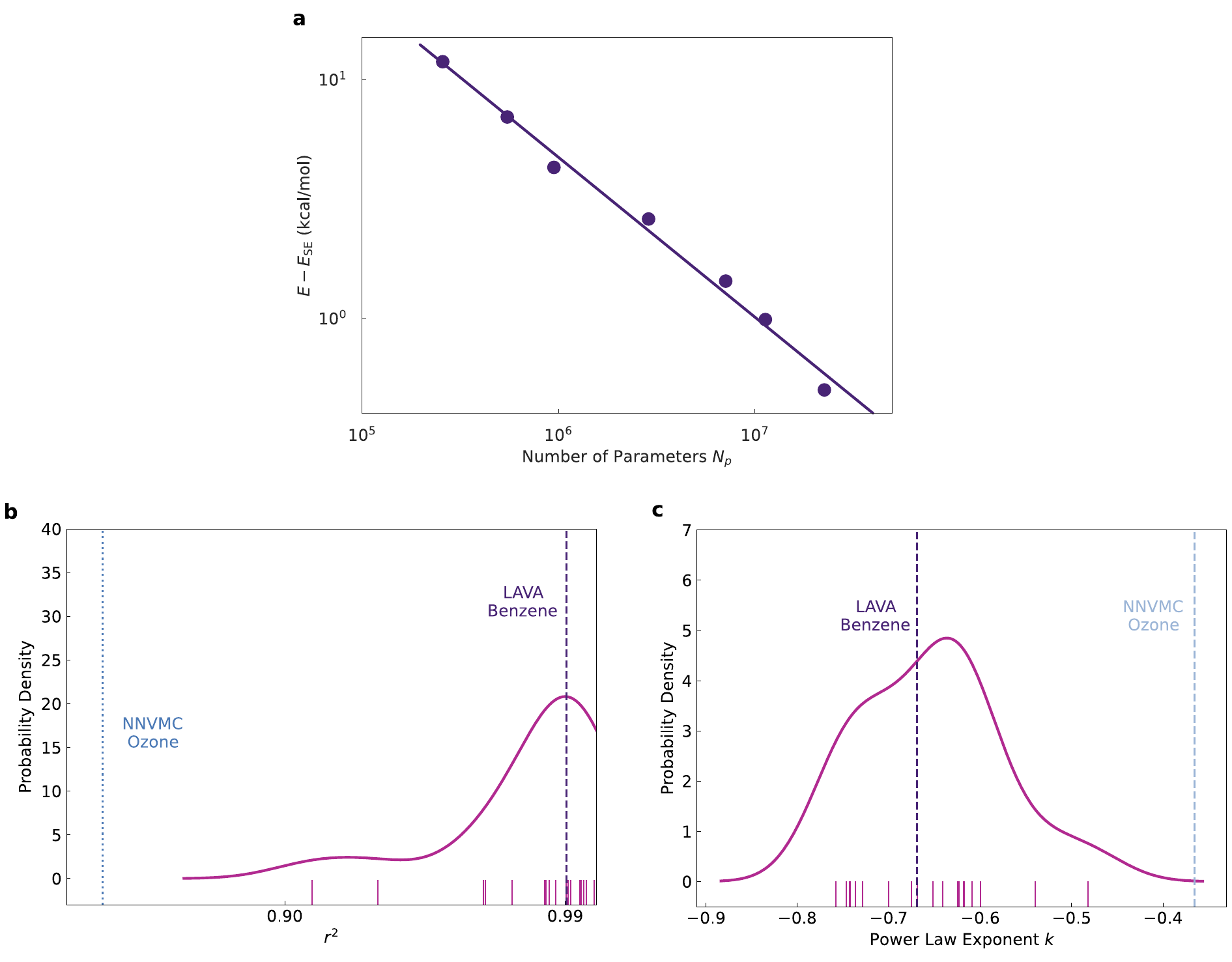}
    \label{fig:stat1}
\end{figure}

\begin{figure}[htp]
    \centering 
    \caption{Power-law scaling trends between local energy variance $\mathrm{Var}[E_L]$ and the number of parameters $N_p$.}
    \includegraphics[width=1.0\textwidth]{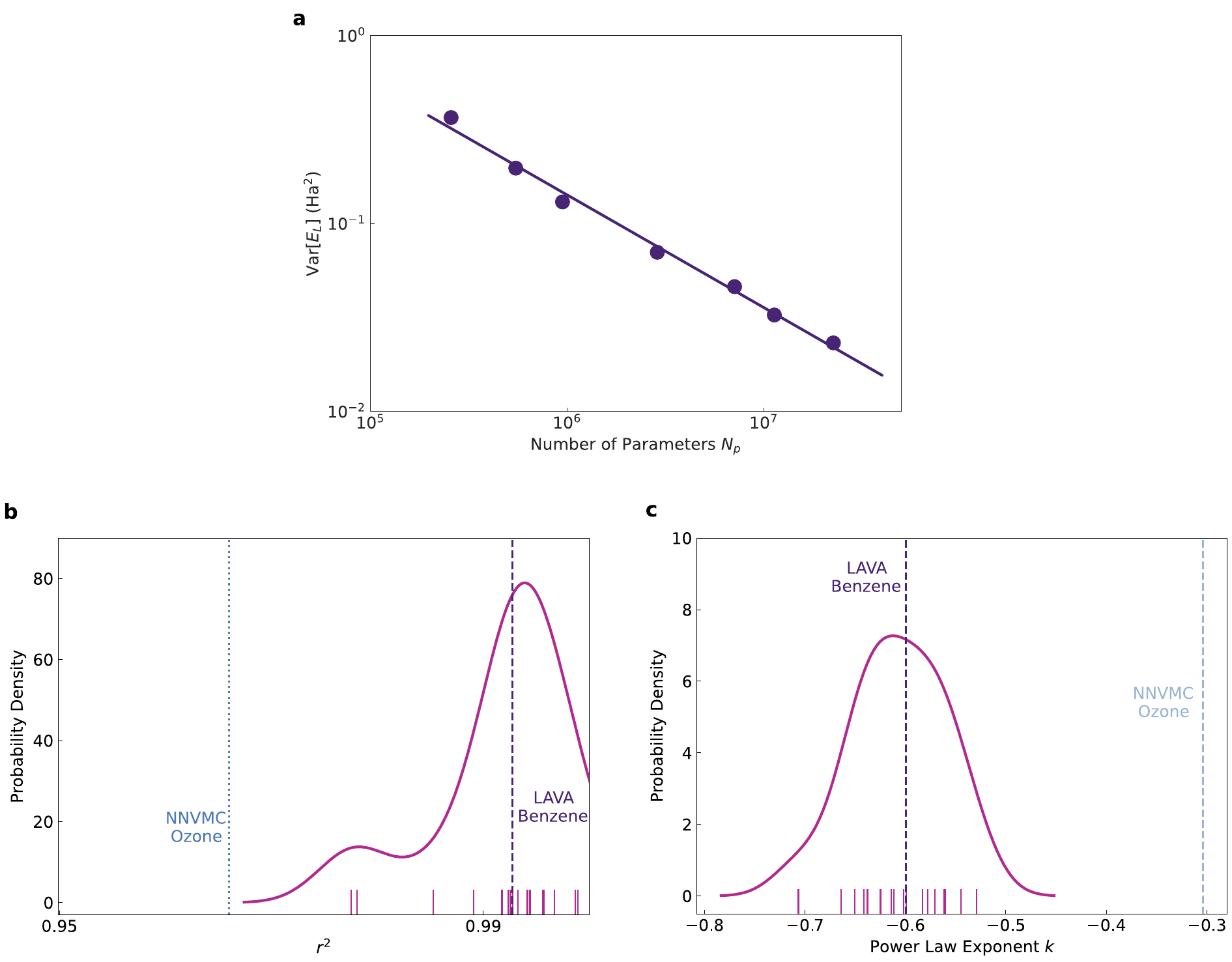}
    \label{fig:stat2}
\end{figure}

As mentioned in Method, we also have general power-law scaling trends between local energy variance $\mathrm{Var}[E_L]$ and the number of parameters $N_p$. For systems without any reliable experimental reference, we can evaluate scaling laws by checking linear relationships between $\ln|E-E_{\mathrm{SE}}|$ or $\ln\mathrm{Var}[E_L]$ with $\ln N_p$. In \Cref{fig:stat1} and \Cref{fig:stat2}, we first take benzene as an example, and afterwards illustrate probability distributions of $r^2$ statistics and power-law exponents for all systems involved in our scaling laws calculations.


\subsection{Comparison against scaling of VMC}

\begin{figure}[htp]
    \centering 
    \caption{LAVA vs. NNVMC scaling-up performance for rectangular cyclobutadiene.}
    \includegraphics[width=1.0\textwidth]{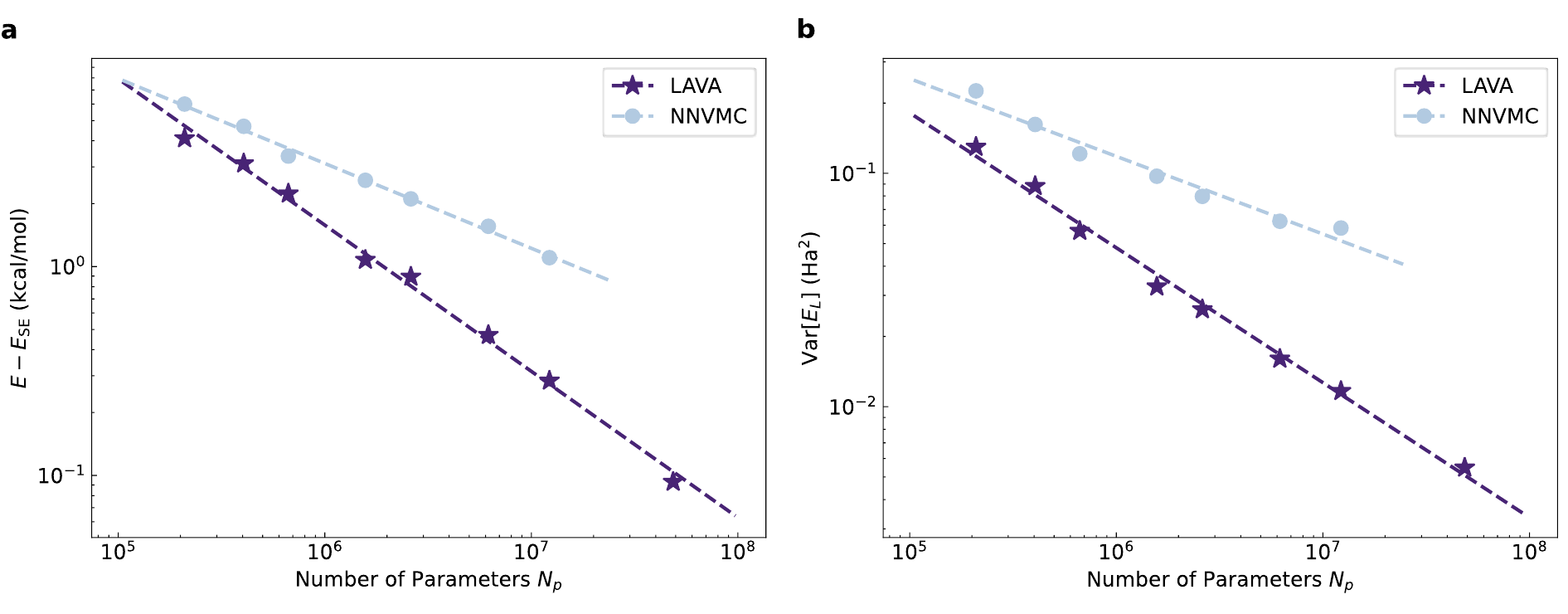}
    \label{fig:fig_scaling_vmc_cycbut}
\end{figure}

\begin{figure}[htp]
    \centering 
    \caption{LAVA vs. NNVMC scaling-up performance for open-minimum ozone.}
    \includegraphics[width=1.0\textwidth]{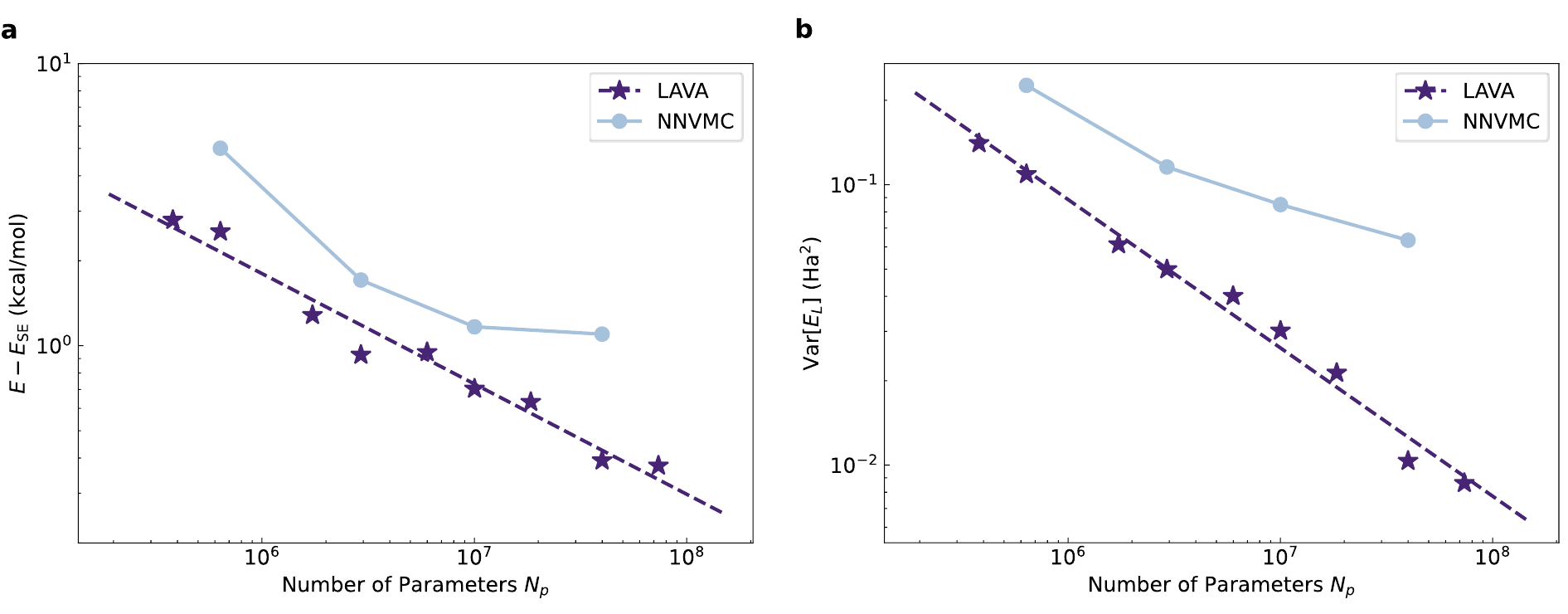}
    \label{fig:fig_scaling_vmc_ozone}
\end{figure}

\Cref{fig:fig_scaling_vmc_cycbut} and \Cref{fig:fig_scaling_vmc_ozone} demonstrate the distinct power-law scaling behavior of LAVA compared to NNVMC. LAVA exhibits clear power-law scaling trends across both test systems, cyclobutadiene and ozone equilibrium, indicating that its accuracy improves systematically and efficiently as computational resources increase. In contrast, NNVMC shows significantly worse power-law scaling on the cyclobutadiene system, and, notably, fails to achieve a good power-law scaling on the ozone equilibrium (open-ring minimum) geometry. This comparison highlights LAVA's superiority over NNVMC in consistent systematic improvability via power-law scaling.

\subsection{Discussion on embarrassingly parallel computing}

LAVA benefits significantly from parallel computing due to its largely independent and homogeneous operations (e.g., matrix multiplications and element-wise functions), which can be efficiently distributed across many processors using data parallelism. In contrast, coupled cluster (CC) methods are not naturally parallelizable and constrained by inherent sequential dependencies in their iterative equations and severe communication overhead for redistributing high-dimensional intermediate tensors. CC's strict synchronization requirements, sensitivity to numerical approximations, and irregular computation patterns fundamentally limit parallel speedup, making scalability beyond modest problem sizes challenging. \citet{Guo2025} recently developed ByteQC, a GPU-accelerated quantum chemistry package featuring CCSD and CCSD(T) modules. In mean-field calculations, electron repulsion integrals (ERIs) with $\mathcal{O}(N^4)$ scaling and tensor contractions between rank-6 tensors in CCSD(T) energy calculations grow rapidly in size with increasing system size $N$,  exceeding CPU/GPU memory capacity. Consequently, data transfers between disk and CPU/GPU memory emerge as a bottleneck for large systems, impeding parallelization.

\section{Extrapolation}
\label{sec:ext}

\begin{figure}[htp]
    \centering 
    \caption{LAVA scaling law extrapolation for benzene.}
    \includegraphics[width=1.0\textwidth]{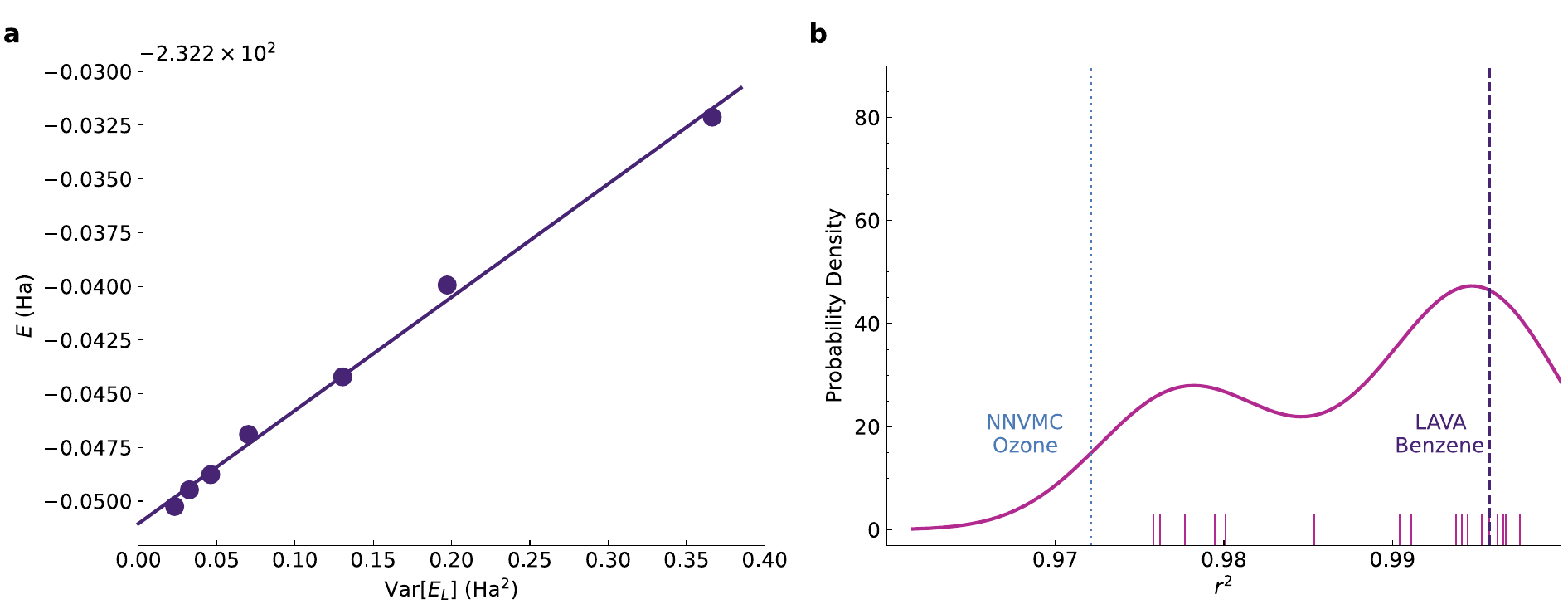}
    \label{fig:fig_scaling_benzene}
\end{figure}

For all LAVA calculations, we universally observe that 
\begin{equation}
E\approx kV+\hat{E}_0,
\end{equation}
given results $E$ and $V$ from optimized models of different scales. The zero-variance principle directly shows that $E_0\approx \hat{E}_0$. We take benzene as an example and illustrate this relationship and corresponding $r^2$ statistics in \Cref{fig:fig_scaling_benzene}. Some similar variance-energy linear relationships in VMC are reported in previous works such as \citet{PhysRevB.48.12037,PhysRevB.58.6800,PhysRevB.91.115106,doi:10.1073/pnas.2122059119,PhysRevB.88.060402,PhysRevB.93.144411}, and \citet{Fu_2024}. Although there is no clear reason for the existence of this linear relationship~\cite{PhysRevB.48.12037,PhysRevB.58.6800}, \citet{Fu_2024} has provided some sufficient and necessary conditions for further understanding. 

\section{Total atomization energy} \label{sec:si tae}

\subsection{How to derive abs. energy benchmark from TAE experiments}
Based on experimental and W4 total atomization energy (TAE) reference value $\Delta E_{\mathrm{TAE}}$ from \citet{karton_w4-11_2011} and absolute atom energies from \citet{Chakravorty1993}, the absolute energy of a molecule is given derived by $\sum_I E_I-\Delta E_{\mathrm{TA}}$, where $\{E_I\}_{I=1}^M$ are absolute atom energies of all $M$ constituent atoms in the molecule. As described by \citet{karton_w4-11_2011}, ATcT \cite{atct} provides TAE$_0$ reference values, namely zero-point inclusive, relativistic, DBOC (Diagonal Born-Oppenheimer Correction) total atomization energies at absolute zero temperature. With thermodynamic correction based on translational, rotational, and vibrational partition functions, one can obtain the TAE at the temperature of experimental conditions or the room temperature 298 K. W4 theory also subtracts the non-electronic contributions from TAE$_0$ and provides zero-point exclusive, non-relativistic, clamped-nuclei TAE$_e$ for direct comparison with electronic structure calculations~\cite{karton_w4-11_2011}.

\subsection{Atomic energies}

For the elements in period 2, LAVA-SE results perfectly agree with the experiment-derived atomic energies \cite{Chakravorty1993} within 0.1 mHa difference, as shown in \Cref{table:atom}. We use the experiment-derived atomic energies to calculate the W2.2, W4, and experiment-derived molecular absolute energies in \Cref{table:atomic_energy}.

\begin{table*}[htbp]
    \centering
    \caption{Atomic energies from various methods in Hartrees.}
    \begin{threeparttable}
    \begin{tabular}{c S[table-format=5.7] S[table-format=5.7] S[table-format=5.7] S[table-format=5.7] S[table-format=5.7]}
    \toprule\midrule
     $\text{Molecule}$ & $\text{DMC}$\cite{Seth2011}$\text{\ (Ha)}$& $\text{FermiNet}$\cite{Pfau2020}$\text{\ (Ha)}$& $\text{NNVMC}$\tnote{a}~$\text{\ (Ha)}$& $\text{LAVA-SE\ (Ha)}$ & $\text{Expt. Derived}$ \cite{Chakravorty1993}$\text{\ (Ha)}$\\
    \midrule
    $\text{C}$ & -37.84446(6) & -37.84471(5) & -37.844811 & \textbf{-37.8449} & -37.8450 \\
    $\text{N}$ & -54.58867(8) & -54.58882(6) & -54.589029 & \textbf{-54.5891} & -54.5892 \\
    $\text{O}$ & -75.0654(1)  & -75.06655(7) & -75.066989 & \textbf{-75.0672} & -75.0673 \\
    $\text{F}$ & -99.7318(1)  & -99.7329(1)  & -99.733559 & \textbf{-99.7339} & -99.7339 \\  
    \midrule\bottomrule
    \end{tabular}
    \label{table:atom}
    \begin{tablenotes}
    \item[a] LapNet is used as the baseline.
    \end{tablenotes}
    \end{threeparttable}
\end{table*}

For the elements in period 3, in contrast, the atomic energies of LAVA's lowest variational energies of the $(4,512,8,128)$ network configuration are lower than those derived from the experiment, as shown in \Cref{table:atom_3rd} and \Cref{fig:fig_si_atoms}. Because LAVA obeys the variational principle, we suggest using LAVA results as the new benchmark.  

\begin{table*}[htbp]
    \centering
    \caption{LAVA's lowest variational energies for third-row atoms.}
        \begin{threeparttable}
    \begin{tabular}{c S[table-format=5.7] S[table-format=5.7] }
    \toprule\midrule
     \text{Molecule}&\text{LAVA}$\text{\ (Ha)}$ & \text{Expt. Derived} \cite{Chakravorty1993}$\text{\ (Ha)}$\\
    \midrule
    $\text{Ne}$ & -128.937943(7)&-128.9376 \\
    $\text{Na}$ &-162.25509(9) &-162.2546 \\
    $\text{Mg}$ &-200.05359(1) &-200.0530 \\
    $\text{Al}$ & -242.34687(2)& -242.3460\\
    $\text{Si}$ &-289.35943(2) &-289.3590 \\
    $\text{P}$ &-341.25925(2) & -341.2590\\
    $\text{S}$ &-398.11087(3) &-398.1100 \\
    $\text{Cl}$ &-460.15045(4) & -460.1480\\
    $\text{Ar}$ &-527.54407(5) & -527.5400\\
    \midrule\bottomrule
    \end{tabular}
    \label{table:atom_3rd}
    \end{threeparttable}
\end{table*}

\subsection{Discrepancy between experimental and theoretical thermochemistry}
\begin{figure}[htp]
    \centering 
    \caption{LAVA's lowest variational energies for studied atoms and molecules.}
    \includegraphics[width=0.75\textwidth]{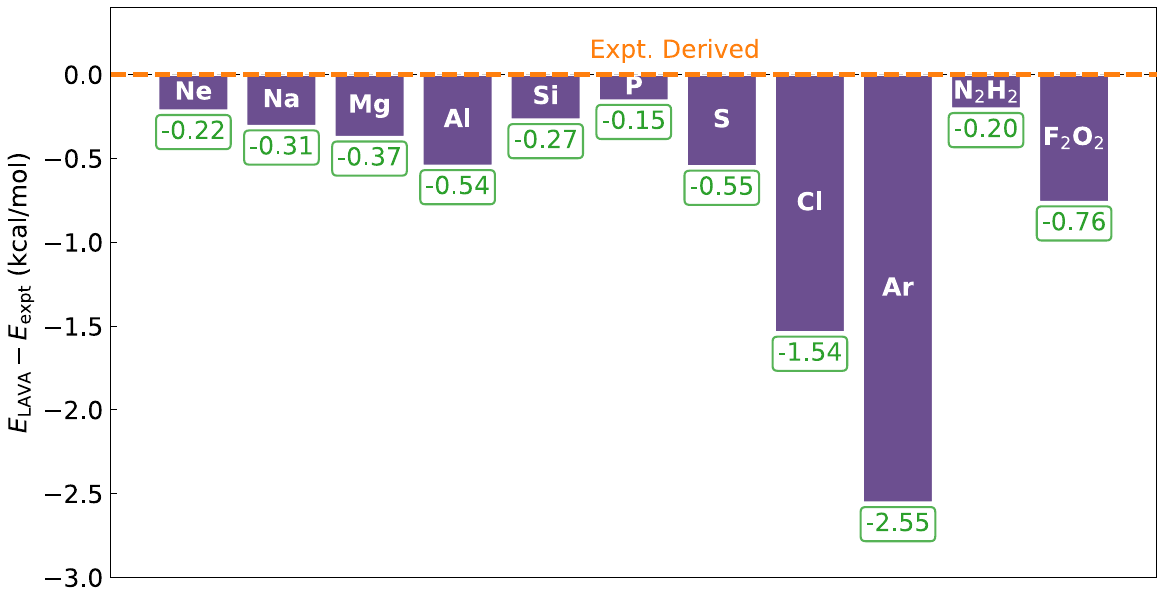}
    \label{fig:fig_si_atoms}
\end{figure}

Since LAVA is variational, it is theoretically guaranteed to give upper bounds of ground state energies. For third-row atoms and certain molecules, we observed that LAVA's best variational energies are lower than experimental-derived reference values, as plotted in \Cref{fig:fig_si_atoms}, which indicates that LAVA's absolute energy results are more reliable for such systems.

In particular, there are several cases where experimental measurements and theoretical prediction have discrepancy, such as $\text{N}_2\text{H}_4$ (hydrazine) \cite{Feller2017}, cis-$\text{N}_2\text{H}_2$ (diazene), and $\text{F}_2\text{O}_2$ (dioxygen difluoride). LAVA shows consensus with composite methods, such as G4 and W4 theory, suggesting the experimental measurements from the 1960s might need re-evaluation. \cite{iftciolu2013, Bond2009}.

Feller \textit{et. al.} proposed the overly optimistic determination of the vaporization enthalpy of  $\text{N}_2\text{H}_4$ to cause the discrepancy between experiments and theories~\cite{Feller2017}.

\subsection{Absolute energies for benzene}
\Cref{table:benzene} provides the results plotted in the main text Fig. 2a. Network configurations are given in 4-tuples that describe network depth, total hidden channels, the number of attention heads, and the number of determinants in sequence.

\begin{table*}[htbp]
    \centering
    \caption{Scaling law results of Benzene. Network configurations are given in the format of (depth, width, the number of attention heads, the number of determinants).}
    \begin{tabular}{c S[table-format=12.7] S[table-format=5.7] S[table-format=5.7] }
    \toprule\midrule
     Network configuration & $N_p$  &\textit{E } \text{\ (Ha)} & \text{Var[$E_L$]} \text{\ (Ha$^2$)} \\
    \midrule
    (4,64,2,16)& 257474& -232.2321(1)&0.3665(2) \\
    (4,96,2,24)& 549026&-232.23994(6) &0.1972(2) \\
    (4,128,4,32)& 949122&-232.24422(6) &0.1304(1) \\
    (4,256,4,32)&2879618 &-232.24689(5) & 0.07033(6) \\
    (4,408,4,48)& 7115954&-232.24877(3) &0.04618(5) \\
    (4,512,8,64)&11329794 &-232.24947(3) &0.03262(3) \\
    (8,512,8,128)& 22633986&-232.25025(2) &0.02318(3) \\
    \midrule
    LAVA-SE &      $\backslash$ & -232.2510  & 0.0 \ \\
    \midrule\bottomrule
    \end{tabular}
    \vspace{-0.3cm}
    \label{table:benzene}
\end{table*}

\subsection{Absolute energies for molecules from W4-11 dataset}
\begin{figure}[htp]
    \centering 
    \caption{Power-law scaling trends for molecules from the W4-11 dataset.}
    \includegraphics[width=1.0\textwidth]{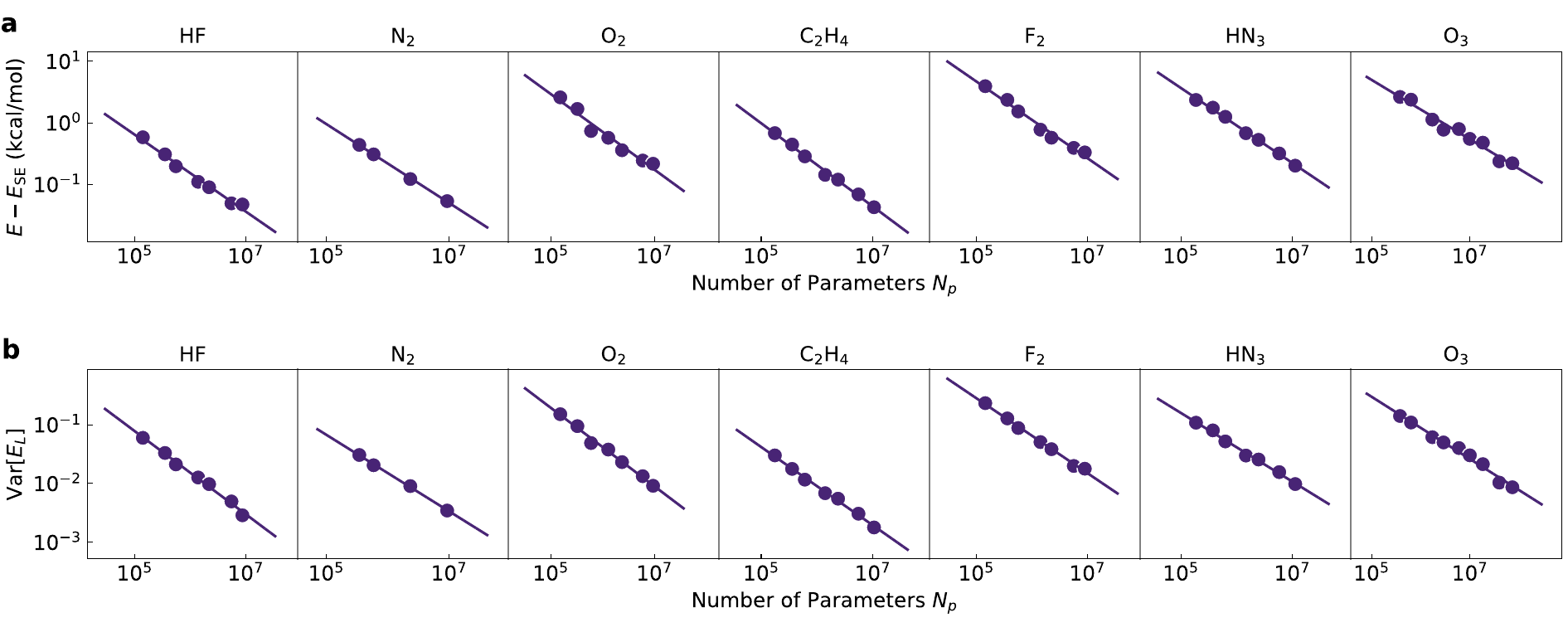}
    \label{fig:scaling_W4}
\end{figure}

\begin{figure}[htp]
    \centering 
    \caption{Variance-energy extrapolation for molecules from W4-11 dataset.}
    \includegraphics[width=1.0\textwidth]{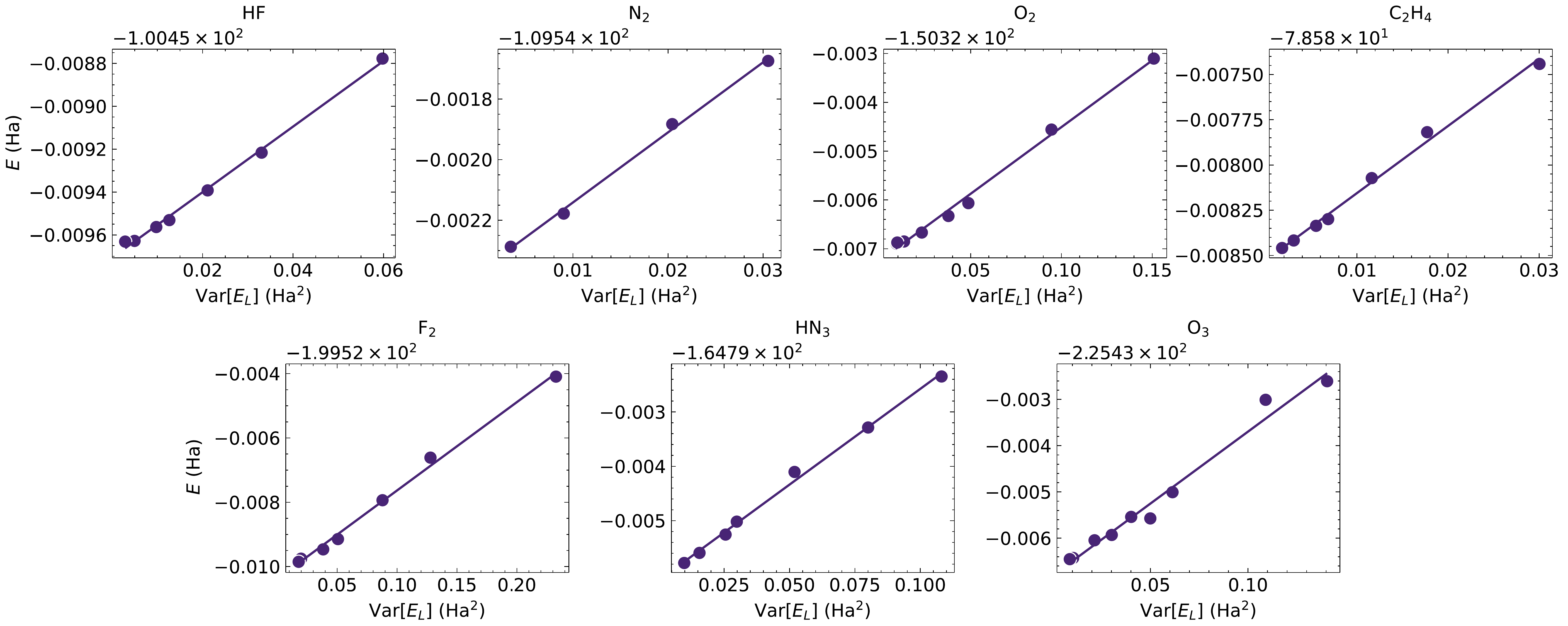}
    \label{fig:scaling_W4_ve}
\end{figure}

\Cref{table:atomic_energy} lists the absolute energies of molecules in the main text Fig. 2b. We use TAE$_e$ values for W2.2 and W4 from W4-11 dataset \cite{karton_w4-11_2011}. The experiment-derived absolute energies are corrected by the non-electronic contributions from W4 theory, TAE$_0$ -TAE$_e$,  to obtain the experimentally derived TAE$_e$ that is directly comparable to LAVA absolute energies. We also illustrate different types of scaling laws in \Cref{fig:scaling_W4} and variance-energy extrapolation in \Cref{fig:scaling_W4_ve}.

\begin{table*}[htbp]
    \centering
    \caption{Absolute energies in Hartrees.}
    \begin{threeparttable}
    \begin{tabular}{lS[table-format=5.7]S[table-format=5.7]S[table-format=5.7]S[table-format=5.7]S[table-format=5.7]}
    \toprule\midrule
     Molecule & {\text{W2.2 Derived}}\text{\ (Ha)}  & \text{W4 Derived}\text{\ (Ha)} & NNVMC\tnote{a}~\text{\ (Ha)}  & \text{LAVA-SE}\text{\ (Ha)}  & \text{Expt. Derived}\text{\ (Ha)}  \\
    \midrule
    $\text{HF}$ &-100.4597 &-100.4597 & -100.4595 & -100.4597& -100.4596\\
    $\text{N}_2$&-109.5419 &-109.5425 & -109.5413 & -109.5424& -109.5424\\
    $\text{O}_2$ & -150.3260&-150.3270 & -150.3245 & -150.3272&-150.3272\\
    $\text{C}_2\text{H}_4$& -78.5889&-78.5889 & -78.5872& -78.5885 & -78.5888 \\
    $\text{F}_2$ & -199.5287 &-199.5299 & -199.5274 & -199.5303&-199.5300\\
    $\text{HN}_3$& -164.7948 & -164.7963 & -164.7916 & -164.7961& -164.7959 \\
    $\text{O}_3$ & -225.4316 & -225.4367 & -255.4304 & -225.4368& -225.4371\\
    \midrule\bottomrule
    \end{tabular}
    \label{table:atomic_energy}
    \begin{tablenotes}
    \item[a] LapNet is used as the baseline.
    \end{tablenotes}
    \end{threeparttable}
\end{table*}

\section{Polyene chains}

\begin{figure}[htp]
    \centering 
    \caption{Power-law scaling trends for polyene chains.}
    \includegraphics[width=1.0\textwidth]{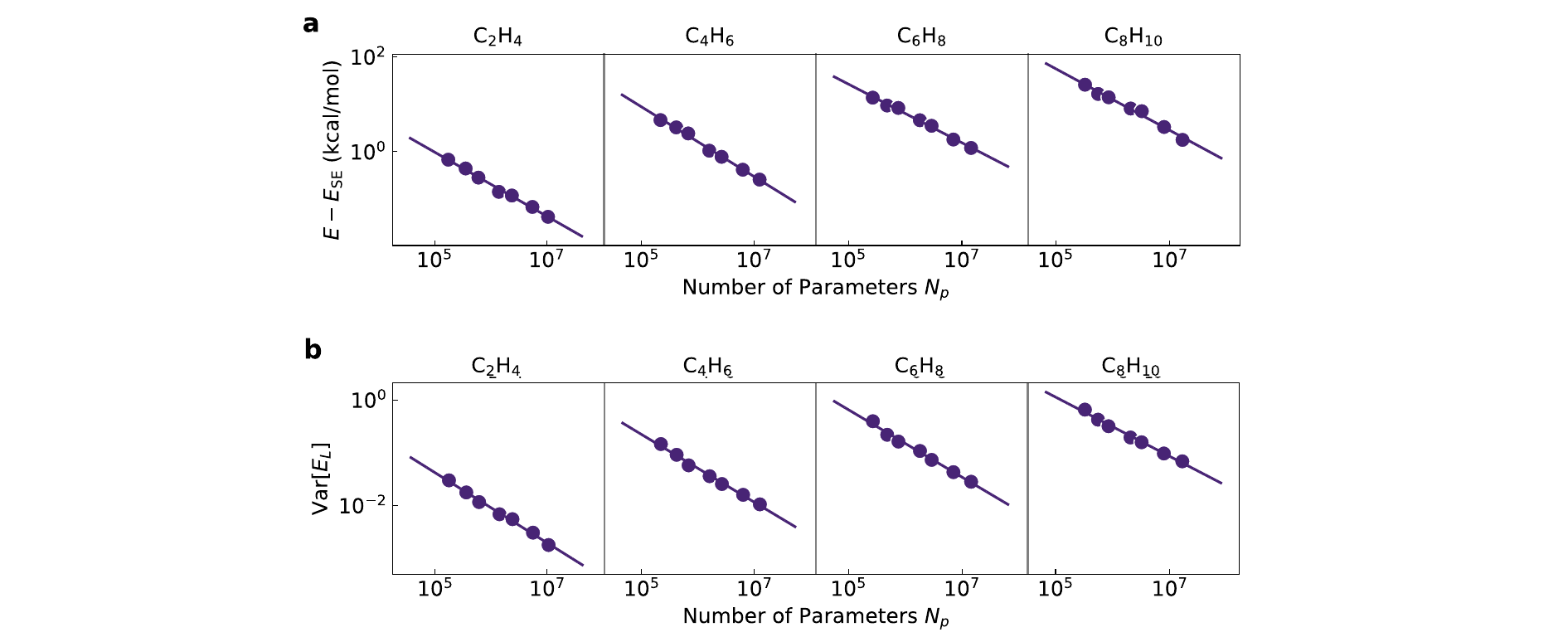}
    \label{fig:scaling_Cn}
\end{figure}

\begin{figure}[htp]
    \centering 
    \caption{Variance-energy extrapolation for polyene chains.}
    \includegraphics[width=1.0\textwidth]{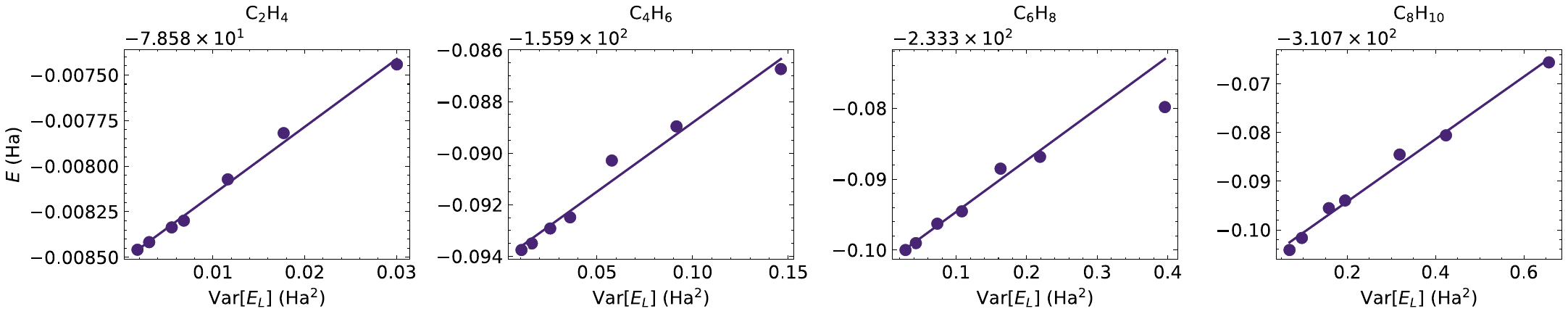}
    \label{fig:scaling_Cn_ve}
\end{figure}

For polyene chains, \Cref{fig:scaling_Cn} illustrates scaling laws and \Cref{fig:scaling_Cn_ve} illustrates variance-energy extrapolation. The neural scaling law $E-E_0=\alpha N_p^{-\beta}$ enables us to evaluate the threshold resources required to achieve a target energy accuracy, e.g., chemical accuracy. In practice, instead of directly inverting the error-parameters power-law relationship for evaluation, we combine variance-parameters power-law scaling $V=\alpha_v N_p^{-\beta_v}$ and extrapolation relationships, which are empirically better supported, to get a revised estimation:
\begin{equation}
N_p^*=\left(\frac{\alpha_v}{V-\Delta V}\right)^{1/\beta_v},
\end{equation}
\begin{equation}
\Delta V=(E-E_{\mathrm{SE}}+\Delta E)/k,
\end{equation}
where $N_p^*$ is the estimated number of parameters required for target accuracy $\Delta E$. Relying on the excellent accuracy of the extrapolation, here we use $E-E_{\mathrm{SE}}$ to estimate the current error. To get a reasonable network setting with $N_p^*$ parameters, we modify the number of network hidden channels $N_C$ and determinants $N_{\mathrm{dets}}$ while maintaining $2048 N_{\mathrm{dets}}=N_C^2$ as default. The GPU hours required to optimize this network can be deemed an approximation of threshold computational costs that avails a priori resource allocation for high-accuracy calculations. To alleviate the influence of multi-host communications, the calculations for time estimation in the main text Fig. 2d were performed on a single $8\times$A800 server.

\section{Cyclobutadiene automerization barrier} 

In this section, we provide a better estimate of the cyclobutadiene automerization barrier. We found that the best estimates of different theoretical predictions reach consensus with each other and also exhibit excellent agreement with our improved estimate of the experimental barrier.

\subsection{Experimental barrier with large uncertainty} \label{sec:si CBD exp}
Whitman \textit{et. al.}  estimated a 1.6-10 kcal/mol automerization barrier based on the trapping reaction between cyclobutadiene and methyl (Z)-3-cyanoacrylate, whose reaction network was given in the Fig. 1 of \citet{Whitman1982}. They measured  $\Delta H^\ddagger_\text{automerization} - \Delta H^{\ddagger}_\text{trapping}=1.6$ kcal/mol. The lower bound of 1.6 kcal/mol was obtained by setting $\Delta H^\ddagger_\text{trapping}$ to zero. The upper limit 10 kcal/mol was estimated by assuming  $\Delta H^\ddagger_\text{trapping}$ is less than $\Delta H^\ddagger=8.3$ kcal/mol, which is measured from cyclopentadiene-benzoquinone Diels-Alder reaction in CCl$_4$ solution.

In order to obtain a better estimate of $\Delta H^\ddagger_\text{trapping}$, we first optimized the structures of reactant, product, and transition state at the B3LYP-D3BJ/aug-cc-pVTZ level. We then calculated the zero-point energy of these structures at the same level of theory. Then, we calculated single-point CCSD(T) energies using cc-pVTZ and cc-pVQZ basis sets and extrapolated the CCSD(T) energies to the complete basis limit, as described in \Cref{sec:SI_CBS_extrapolation}.  
The resulting $\Delta H^\ddagger_\text{automerization}$ is 9.9 kcal/mol, which is in good agreement with the estimated upper limit in \citet{Whitman1982}.

\subsection{Complete basis set extrapolation}
\label{sec:SI_CBS_extrapolation}
Therefore, we follow the two-point extrapolation scheme proposed in \citet{halkier1999} for Dunning’s correlation-consistent series of Gaussian basis sets. The extrapolated HF energy is written as
\begin{equation}
    E^{\rm{HF}}_{\infty} = E^{\rm{HF}}_{n} - \frac{E^{\rm{HF}}_{n}-E^{\rm{HF}}_{n+1}}{1-e^{-B}} ,
\end{equation}
where $B$ is a constant number 1.637 and $n$ is the $\zeta$ cardinality for basis set.

The extrapolated correlation energy follow the formula~\cite{halkier1998}
\begin{equation}
    E^{\rm{corr}}_{\infty} = \frac{n^3 E^{\rm{corr}}_n - m^3 E^{\rm{corr}}_m}{n^3 - m^3} ,
\end{equation}
where $n$ and $m$ are the $\zeta$ cardinality for basis sets, usually $n=m+1$.

\subsection{Available experimental observables of cyclobutadiene}

We further calculated total atomization energy, heat of formation, and ionization potential of cyclobutadiene, where experimental measurements are available. LAVA shows perfect agreement with experiments.  

LAVA predicts total atomization energy of 820.46 kcal/mol, which is in good agreement with 820.38 and 820.72  kcal/mol at the  CCSD(T) and CCSDT(Q)level. \cite{Bakowies2019}.

In addition, LAVA predicts an heat of formation $\Delta_fH_0^\circ(\text{C}_4\text{H}_4)=$ 104.7 kcal/mol for $2\text{C}_2\text{H}_2 \rightarrow \text{C}_4\text{H}_4$, where the zero point energy correction 4.83  kcal.mol/mol at CCSD(T)/cc-pVTZ level is taken from \citet{Wu2012}. This is good agreement with CCSDT(Q)/CBS prediction of 104.2 $\pm$ 1.0 kcal/mol\cite{Wu2012}  and experimental values of 102.3  $\pm$ 3.8 kcal/mol.\cite{Fattahi2006}

Finally, LAVA-SE predicts an ionization potential of 8.165 eV, which is in excellent agreement with the experimental measurement of 8.16$\pm$0.03 eV\cite{Kohn1993}. In contrast, the theoretical value from various CC-based composite methods reported by ATcT (8.023–8.078 eV)\cite{atct} lies outside the experimental uncertainty range.

\begin{figure}[htbp]
    \centering 
    \caption{Power-law scaling trends for cyclobutadiene. From left to right, the absolute energy of rectangular cyclobutadiene, the absolute energy of square cyclobutadiene, and the first ionization potential of rectangular cyclobutadiene.}
    \includegraphics[width=1.0\textwidth]{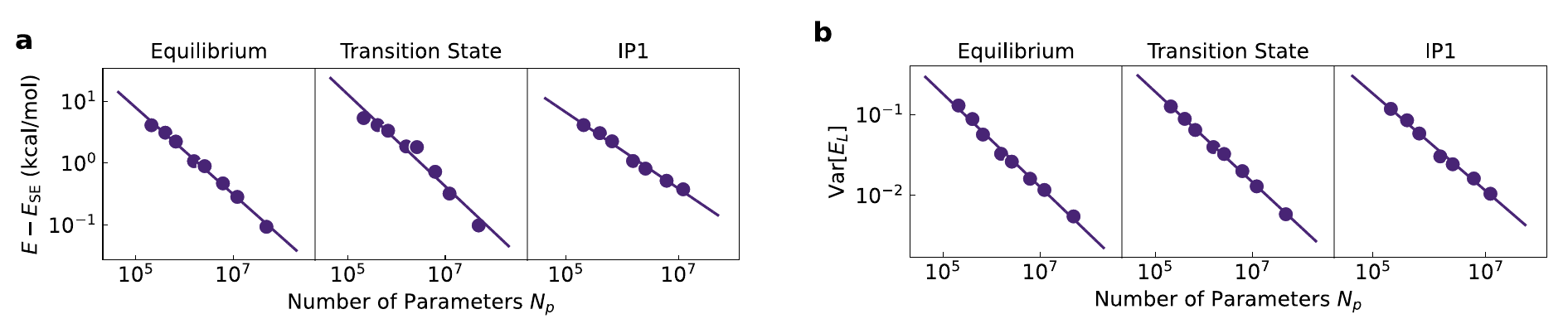}
    \label{fig:scaling_cycbut}
\end{figure}

\begin{figure}[htbp]
    \centering 
    \caption{Variance-energy extrapolation for cyclobutadiene. From left to right, the absolute energy of rectangular cyclobutadiene, the absolute energy of square cyclobutadiene, and the first ionization potential of rectangular cyclobutadiene.}
    \includegraphics[width=1.0\textwidth]{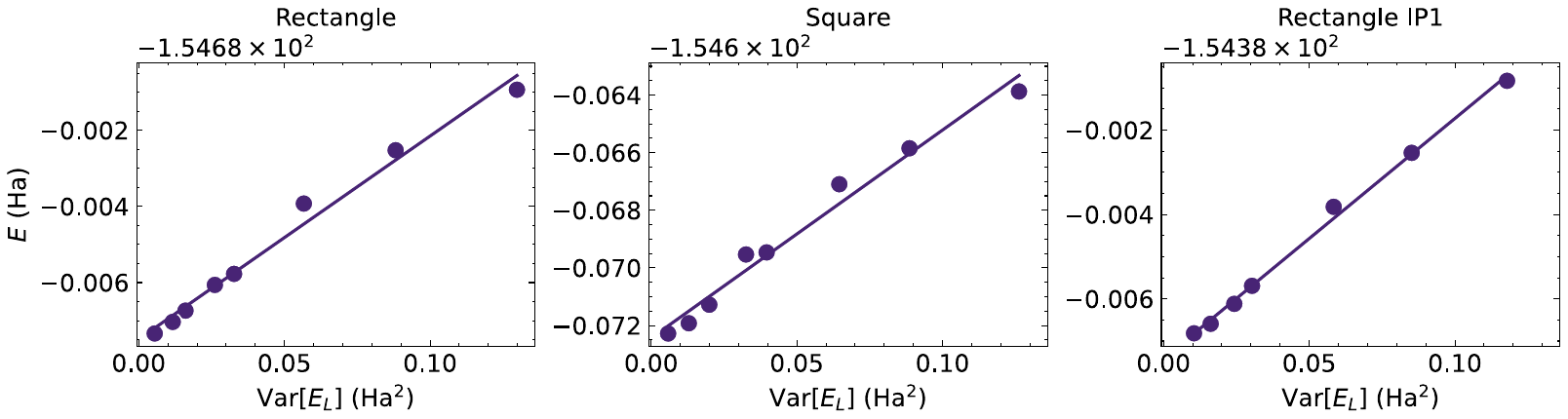}
    \label{fig:scaling_cycbut_ve}
\end{figure}

\Cref{tab:CBD} provides the LAVA results plotted in the main text Fig. 3a. Network configurations are given in 4-tuples that describe network depth, total hidden channels, the number of attention heads, and the number of determinants in sequence. \Cref{fig:scaling_cycbut} illustrates scaling laws and \Cref{fig:scaling_cycbut_ve} shows the variance-energy extrapolation.

\begin{table}[htbp]
\centering
\caption{Absolute energies, first ionization potential, and automerization barrier of cyclobutadiene from LAVA. Network configurations are given in the format of (depth, width, the number of attention heads, the number of determinants).}
\label{tab:CBD}
\sisetup{separate-uncertainty, table-format = -1.6} 
\begin{tabular}{lS[table-format=5.7]S[table-format=5.7]S[table-format=5.7]S[table-format=5.7]S[table-format=5.7]} 
\toprule\midrule
Network configuration & \text{Rectangular\ (Ha)} & \text{Rectangular\ anion\ (Ha)} & \text{Square\ (Ha)} &  \text{IP1\ (eV)} & \text{Barrier\ (kcal/mol)}\\
\midrule
(4,64,2,16)    &-154.68093(6)& -154.38083(6)&-154.66387(6)  &8.166 &10.70\\
(4,96,2,16)    &-154.68252(5)& -154.38254(5)&-154.66585(5) &8.163 &10.46\\
(4,128,4,16)   &-154.68392(4)&-154.38382(4) &-154.6671(4) &8.166 &10.56\\
(4,192,4,32)   &-154.68577(3)&-154.38569(3) &-154.66946(3) &8.166 &10.23 \\
(4,256,4,32)   &-154.68606(3)&-154.38612(3) &-154.66953(3) &8.162 &10.37\\
(4,384,4,64)   &-154.68674(2)&-154.38659(2) &-154.67128(2) &8.167 &9.70  \\
(4,512,8,128)  &-154.68703(2)&-154.38682(2) &-154.67192(2) &8.169 &9.48   \\
(4,1024,16,256)&-154.68734(1) & \textbackslash &-154.67228(1) & \textbackslash    &9.45 \\
\midrule
LAVA-SE      &-154.687487 &-154.387417 &-154.672434 & \textbf{8.165}& \textbf{9.44}\\
\midrule\bottomrule
\end{tabular}
\end{table}

Notably, the first ionization potential (IP1) remains accurate even with the smallest network (4,64,2,16), exhibiting only minor fluctuations as network size increases. In contrast, the automerization barrier shows initial fluctuations but decreases significantly when scaling from network size (4,256,4,32) to (4,384,4,64). 

\subsection{Comparison of cyclobutadiene automerization barrier from various theoretical methods }

\begin{table}[htbp]
\centering
\caption{Reaction barrier of cyclobutadiene in kcal/mol from references.}
\label{tab:CBD_barrier_ref}
\sisetup{separate-uncertainty, table-format = -1.6} 
\begin{tabular}{lrrrrr} 
\toprule\midrule
CC\cite{Monino2022} & 6-31+G(d)  & aug-cc-pVDZ   & aug-cc-pVTZ  & aug-cc-pVQZ  \\
\midrule
CCSD            & 8.31       & 8.80          & 9.88         & 10.10  \\
CC3             & 6.59       & 6.89          & 7.88         & 8.06  \\
CCSDT           & 7.26       & 7.64          & 8.68         & 8.86  \\
CC4             & 7.40       & 7.78          & 8.82         & 9.00  \\
CCSDTQ          & 7.51       & 7.89          & 8.93         & \textbf{9.11}  \\
\midrule
CC\cite{Lyakh2011} &          & cc-pVDZ       & cc-pVTZ  &  \\
\midrule
CCSD            &            & 21.0          & 23.2         &       \\
CCSD(T)         &            & 15.8          & 18.3         &       \\
CR-CCSD(T)      &            & 18.3          & \textbackslash  &       \\
$\Lambda$CCSD(T) &           & 16.8          & 19.2         &       \\
\midrule
MRPT\cite{Monino2022} & 6-31+G(d)  & aug-cc-pVDZ   & aug-cc-pVTZ  & aug-cc-pVQZ  \\
\midrule
CASPT2(4,4)     & 6.56       & 6.87          & 7.77         & 7.93  \\
SC-NEVPT2(4,4)  & 7.95       & 8.31          & 9.23         & 9.42  \\
PC-NEVPT2(4,4)  & 7.95       & 8.33          & 9.24         & 9.41  \\
CASPT2(12,12)   & 7.24       & 7.53          & 8.51         & 8.71  \\
SC-NEVPT2(12,12)& 7.10       & 7.32          & 8.29         & 8.51  \\
SC-NEVPT2(12,12)& 7.12       & 7.33          & 8.28         & 8.49  \\
\midrule
MRCC\cite{Lyakh2011} &          & cc-pVDZ       & cc-pVTZ   &       \\
\midrule
TCCSD              &            & 9.4           & 12.9      &       \\
SUCCSD             &            & 7.0           & 8.7       &       \\
BWUCCSD(a.c)       &            & 6.5           & 7.6       &       \\
BWUCCSD(i.c)       &            & 6.2           & 7.4       &       \\
MkCCSD             &            & 7.8           & 9.1       &       \\
RMRCCSD            &            & 10.4          & 13.0      &       \\
TCCSD(T)           &            & 4.6           & 7.0       &       \\
SUCCSD(T)          &            & 4.8           & 5.9       &       \\
BWUCCSD(T)(a.c)    &            & 6.1           & 7.0       &       \\
BWUCCSD(T)(i.c)    &            & 5.7           & 6.8       &       \\
MkCCSD(T)          &            & 7.8           & 8.9       &       \\
RMRCCSD(T)         &            & 7.2           & 9.5       &       \\
\midrule
Select CI          &            & cc-pVDZ       & cc-pVTZ   &       \\
\midrule
iFCI \cite{Hatch2025}  &        & 7.55          & 8.43       &       \\
FCI 4-point extrapolation \cite{Dang2022}    &        & \             & \textbf{9.21}       &       \\
\midrule
NNQMC         &            &       &    &       \\
\midrule
PauliNet\cite{hermann_deep-neural-network_2020}      & \multicolumn{2}{c}{9.9 $\pm$ 0.6 (250 steps)}  & \multicolumn{2}{c}{7.7 $\pm$ 0.6 (375 steps)} \\
FermiNet-VMC\cite{arxiv.2011.07125}  &  \multicolumn{2}{c}{10.3 $\pm$ 0.1 ($2 \times 10^5$ steps)} & \\
FermiNet-VMC\cite{Ren2023}  & \multicolumn{2}{c}{9.22($1 \times 10^5$ steps)}   & \multicolumn{2}{c}{9.41 ($5 \times 10^5$ steps)}  \\
FermiNet-DMC\cite{Ren2023}   & \multicolumn{2}{c}{9.73($1 \times 10^5$ steps)}  & \multicolumn{2}{c}{9.98 ($5 \times 10^5$ steps)} \\

\midrule\bottomrule
\end{tabular}
\end{table}

Traditional quantum chemistry methods based on second quantization suffer from inherent difficulty in describing the multireferential character of the transition state, while experimentalists can hardly capture the highly unstable transition states. Previous calculations, as listed in \Cref{tab:CBD_barrier_ref}, have not achieved a consensus. The perfect agreement between LAVA, the best estimates of systematically improvable methods, and improved experiments suggests that LAVA can serve as the benchmark for studying transition states and reaction kinetics.

\section{New benchmark for N\textsubscript{2} potential energy curve}\label{sec:si n2}

The potential energy curve (PEC) of N$_2$ is widely studied by various theoretical methods. The most commonly used benchmarks are theoretical r$_{12}$-MR-ACPF of \citet{Gdanitz1998} and analytic PECs fitted from experimental spectroscopic data~\cite{LeRoy2006}. \citet{LeRoy2006} fitted the ground-state PEC based on vibrational levels up to $v$=19, covering bond lengths $r$ between 0.9 and 1.5 $\text{\AA}$.\cite{LeRoy2006} At the fully dissociation limit $r> 4\ \text{\AA}$, the behavior of the analytic fitting function is governed by the experimentally measured dissociation energy $D_e$ and Morse long-range (MLR) function. However, these analytic potential functions show large uncertainty within the nearly dissociation region $2\ \text{\AA}<r < 4\ \text{\AA}$ as shown in the Fig. 2 of \citet{LeRoy2006}.  At $r=2.2\ \text{\AA}$, the discrepancy between  $\text{MLR}_4(6,8)$ and $\text{EMO}_2(6)$ is 1210 cm$^{-1}$, namely 3.4 kcal/mol, as shown in the inset of our main text Fig. 3b. Within the gray shaded region, where experimental vibrational levels are available, LAVA perfectly aligns with  $\text{MLR}_4(6,8)$. At the stretched bond length ($r=1.5-3\ \text{\AA}$), the certainty of LAVA remains the same as the equilibrium bond length, suggesting the reliability of LAVA across the entire PEC. The previous theoretical SOTA r$_{12}$-MR-ACPF perfectly agree with LAVA but shift up by about 6 mHa due to the finite basis set limitation. Since LAVA is based on first quantization, it is not limited by a finite basis set as the second-quantized methods are. Therefore, LAVA might have advantages over second-quantized methods in chemical systems where the size of the finite basis set plays an important role, such as transition metal complexes, excited states\cite{Pfau2024}, and positronic chemistry~\cite{Cassella2024}. 

First, we fitted various EMO and MLR curves using the 1221 experimental data points from \citet{LeRoy2006} and software dPotFit \cite{LeRoy2017}. Then, we calculated the vibrational energy levels of these fitted curves using software LEVEL \cite{LeRoy2017level}. Finally, we chose the fitted curve that agrees best with our LAVA prediction, especially within the nearly dissociation region. Therefore, we retain the high accuracy of the original $\text{MLR}_4(6,8)$ curve around the equilibrium region between 0.9 and 1.5 $\text{\AA}$ while improving the reliability at the nearly dissociation region between 1.8 and 3.0 $\text{\AA}$.

\subsection{Solving effective radial Schrödinger equation}
The vibrational $v$ and rotational $J$ energy levels can be obtained by solving the effective radial Schrödinger equation:

\begin{equation}
\Biggr\{-\frac{\hbar^2}{2\mu}  \frac{d^2}{dr^2} + V(r)+\frac{\hbar^2 J(J+1)}{2\mu r^2}   \Biggr\} \psi_{v,J}(r) = E_{v,J}\psi_{v,J}(r)
\label{eq:V}
\end{equation}

where $V$ is the total effective adiabatic internuclear potential, $\mu$ is the reduced mass of two atoms forming N$_2$;  the last term on the left-hand-side $\frac{\hbar^2 J(J+1)}{2\mu r^2}  $ is called the nonadiabatic centrifugal term.

The reference isotopologue is $^{14,14}\text{N}_2$, labled by $\alpha=1$. For the other isotopologues $^{14,15}\text{N}_2$ and $^{15,15}\text{N}_2$, \citet{LeRoy2006} used adiabatic correction term $\Delta V^\alpha$ and nonadiabatic correction term $g^\alpha$ for isotopologue $\alpha$.

\begin{equation}
\Biggr\{-\frac{\hbar^2}{2\mu_\alpha}  \frac{d^2}{dr^2} + \bigl[V^{(1)}(r)+\Delta V^{(\alpha})(r)\bigr] + \frac{\hbar^2 J(J+1)}{2\mu_\alpha r^2}  \bigl[ 1+ g^{(\alpha)}(r)\bigr] \Biggr\} \psi_{v,J}(r) = E_{v,J}\psi_{v,J}(r)
\label{eq:V_isotope}
\end{equation}

\subsection{Fitting analytic potential function $V(r)$ }
\citet{LeRoy2006} used 1221 spectroscopic data points of  $^{14,14}\text{N}_2$ , $^{14,15}\text{N}_2$, and $^{15,15}\text{N}_2$ from Raman and electric quadrupole vibration-rotation experiments of $^{14,14}\text{N}_2$, $^{14,15}\text{N}_2$, and $^{15,15}\text{N}_2$. One can obtain the theoretical prediction of vibrational $v$ and rotational $J$ data by plugging the analytic potential function $V(r)$ into equations \ref{eq:V} and \ref{eq:V_isotope}, and then comparing the experimental and theoretical results. The quality of the analytic potential function fit is defined by the dimensionless root mean square deviation $\overline{dd}$:
 \begin{equation}
\overline{dd} \equiv \Biggl\{\frac{1}{N} \sum^N_{i=1} \biggl[\frac{y_\text{calc}(i)-y_\text{obs}(i)}{u(i)} \biggr]^2 \Biggr\}^{1/2}
\end{equation}

where $N$ is the number of experimental data, $u(i)$ is the estimated experimental uncertainties given in Table I of \citet{LeRoy2006}, $y_\text{calc}(i)$  and $y_\text{obs}(i)$ are the calculated and experimental values, respectively. $\overline{dd}$ value reflects the difference between prediction and experiments is $\overline{dd}$ times of the estimated experimental uncertainties. The smaller the $\overline{dd}$ value, the better the fit.

\subsection{Expanded Morse oscillator function $V_\text{EMO}(r)$}
One expression of the $V(r)$ term in eq \ref{eq:V} is “expanded Morse oscillator” (EMO) function:
\begin{equation}
V_\text{EMO}(r)=\mathcal{D}_e[1-e^{-\phi(r)(r-r_e)}]^2
\label{eq:EMO}
\end{equation}
where $\mathcal{D}_e$ is the dissociation energy, $r_e$ is the equilibrium distance, and $\phi(r)$ is a power series expansion,
\begin{equation}
\phi(r)=\phi_\text{EMO}(r)=\sum^N_{i=0}\phi_i y_p (r)^i 
\label{eq:EMO_N}
\end{equation}
and
\begin{equation}
y_p (r)=\frac{r^p-r_e^p}{r^p+r_e^p}
\label{eq:EMO_p}
\end{equation}

EMO$_2$(6) is the upper bound of various fitted functions in Fig. 2 of \cite{LeRoy2006} as well as Fig. 3b of this work.  EMO$_2$(6) uses $N=6$ in eq.\ref{eq:EMO_N} and  $p=2$ in eq.\ref{eq:EMO_p}.  The coefficients $\phi_0$ to $\phi_6$  are listed in \Cref{tab:MLR parameters}. 

\subsection{Morse/long-range potential function $V_\text{MLR}(r)$}

$V_\text{EMO}(r)$ decays exponentially at large $r$, while the realistic PEC decays in inverse power at the long range. To incorporate the correct asymptote behavior, \citet{LeRoy2006} proposed the Morse/long-range (MLR) potential form:

\begin{equation}
V_\text{MLR}(r)=\mathcal{D}_e \Biggl\{1-(\frac{r_e}{r})^n \biggl[\frac{1+R_\text{m,n}/r^{m-n}}{1+R_\text{m,n}/r_e^{m-n}}\biggr] e^{-\phi(r)y_p(r)} \Biggr\}^2
\label{eq:MLR}
\end{equation}

\begin{equation}
\phi(r)=\phi_\text{MLR}(r)=[1-y_p(r)]\sum_{i=0}^N \phi_iy_p(r)^i+y_p(r)\phi_\infty
\end{equation}

\begin{equation}
\phi_\infty = \text{ln}\frac{2\mathcal{D}_e(r_e)^n}{C_n[1+R_{m,n}/r_e^{m-n}]}
\end{equation}

MLR function in eq. \ref{eq:MLR} resembles the Morse function form in eq. \ref{eq:EMO}. Unlike the EMO function, the MLR function is simplified to two long-range inverse-power terms  $V_\text{MLR}(r) \approx\mathcal{D}_e -\frac{C_6}{r^6} - \frac{C_8}{r^8} $ when $r \rightarrow \infty$ when $n=6$ and $m=8$. Therefore, $C_n$, $R_{m,n}$, $\mathcal{D}_e$, and $r_e$ determine the asymptotic behavior of MLR function, together with $\phi_i$ and $p$ values affect the shape of intermediate distances.

\citet{LeRoy2006} recommended MLR$_4$(6,8) as the best fit, indicating $p=4$, $N_S=6$, and $N_L=8$ in the following equations: 
\begin{equation}
\phi_\text{EMO}(r)=\sum^{N_S}_{i=0}\phi_i y_p (r)^i  \text{ for } r \leq r_e 
\end{equation}

\begin{equation}
\phi_\text{EMO}(r)=\sum^{N_L}_{i=0}\phi_i y_p (r)^i  \text{ for  } r > r_e 
\end{equation}

The reasons for using mixed exponent polynomial orders ($N_s \neq N_L$) were discussed in detail in \citet{LeRoy2006}. In the later works of the Le Roy group, they used $N_s= N_L$ to avoid discontinuity at $r = r_e$. Following the notation in \citet{LeRoy2006}, we label the models by $\text{MLR}_p(N)$ or EMO$_p(N)$ when $N_s=N_L$, and $\text{MLR}_p(N_S, N_L)$ when $N_s \neq  N_L$.  $\text{MLR}_3(9)$ refers to an MLR potential with the same exponent polynomial of order $N_s=N_L=N=9$ together with the expansion variable $y_3(r)=\frac{r^3-r^3_e}{r^3+r^3_e}$. Small $p$ value leads to a gradual change of PEC around $r=r_e$ while a large $p$ value indicates a sharper change. Parameters to define MLR$_3$(9) and MLR$_4$(6,8) curves and the difference $\overline{dd}$ between the experimental and calculated vibrational levels are listed in \Cref{tab:MLR parameters}.

\begin{table}[H]
\centering
\caption{Parameters to define the MLR$_4$(6,8), MLR$_3$(9), and  EMO$_2$(6) curves for the $X^1\Sigma^+_g$ state of $\text{N}_2$.}
\label{tab:MLR parameters}
\sisetup{separate-uncertainty, table-format = -1.6} 
\begin{tabular}{l S[table-format=1.8] S[table-format=1.8] S[table-format=1.8]} 
\toprule\midrule
\textbf{Potential function} & \textbf{MLR$_4$(6,8)} & \textbf{MLR$_3$(9)} & \textbf{EMO$_2$(6)}\\ 
\midrule
$\mathcal{D}_e$/cm$^{-1}$        & \multicolumn{3}{c}{79845} \\
$r_e/\text{\AA}$                 & \multicolumn{3}{c}{1.097679} \\
\midrule
$C_6/(\text{cm}^{-1}\cdot\text{\AA}^6)$ & \multicolumn{2}{c}{1.16 $\times 10^5$} & $\backslash$ \\
$R_{8,6}/\text{\AA}^2$ & \multicolumn{2}{c}{5.5} & $\backslash$\\
\midrule
$\phi_0$                & -2.34414547        &  3.125518   & 2.689598 \\
$\phi_1$                & -0.972469          & -1.338651   & 0.332343 \\
$\phi_2$                & -1.561777          & -1.926378   & 0.665061 \\
$\phi_3$                & -1.136             & -1.053483   & 0.729573 \\
$\phi_4$                & -1.3963            & -1.088594   & 1.157254 \\
$\phi_5$                & -0.819             &  0.253611   & 2.824968 \\
$\phi_6$                & -0.45              &  0.767715   &12.341291 \\
$\phi_7$                & -3.36              & -1.929323   & $\backslash$       \\
$\phi_8$                & 2.1                &  1.583792   & $\backslash$       \\
$\phi_9$                & $\backslash$               & 15.210659   & $\backslash$       \\
\midrule
\addlinespace[4pt]
$\overline{dd}$           & 1.44               & 1.42         & 1.43 \\ 
\midrule\bottomrule
\end{tabular}
\end{table}

\subsection{Low-lying vibrational levels for $v$=0-19}
$v$=0-19 are included in 1221 experimental data points when fitting analytic potential functions. In \Cref{tab:G and B} , we listed the vibrational levels $G_v-G_0$  and rotational constants $B$  of experiment, MLR$_4$(6,8), and MLR$_3$(9). The new benchmark MLR$_3$(9) gives slightly smaller   RMSD in vibrational levels and the small RMSD in rotational constants than the old benchmark MLR$_4$(6,8), suggesting MLR$_3$(9) is superior to MLR$_4$(6,8) in reproducing experimental results.

\begin{table}[htbp]
\centering
\caption{Vibrational levels $G_v-G_0$ and rotational constants $B$ from experiments for the $X^1\Sigma^+_g$ state of $\text{N}_2$, and the errors of MLR$_4$(6,8) and MLR$_3$(9) curves compared to experiments\cite{Laher1991}. }
\label{tab:G and B}
\sisetup{separate-uncertainty, table-format = -1.6} 
\begin{tabular}{lrrrrrr} 
\toprule\midrule
\multirow{2}{*}{\makecell{Level}} &  \multicolumn{3}{c}{$G_v-G_0(\text{cm}^{-1}$)} & \multicolumn{3}{c}{$B$(cm$^{-1}$)} \\
\cmidrule(lr){2-4} \cmidrule(lr){5-7}
 & \text{Experiment} & \text{MLR$_4$(6,8)} & \text{MLR$_3$(9)} & \text{Experiment} & \text{MLR$_4$(6,8)} & \text{MLR$_3$(9)} \\
\midrule
0    & 0.00      & 0.00   & 0.00 & 1.9896 &  0.0000 &  0.0000 \\
1    & 2329.91   & 0.00   & 0.00 & 1.9722 &  0.0000 &  0.0000 \\
2    & 4631.17   & 0.01   & 0.02 & 1.9547 &  0.0001 &  0.0001 \\
3    & 6903.72   & 0.03   & 0.04 & 1.9372 &  0.0001 &  0.0001 \\
4    & 9147.54   & 0.05   & 0.06 & 1.9196 &  0.0001 &  0.0001 \\
5    & 11362.59  & 0.05   & 0.07 & 1.9020 &  0.0002 &  0.0002 \\
6    & 13548.81  & 0.05   & 0.07 & 1.8843 &  0.0003 &  0.0003 \\
7    & 15706.18  & 0.04   & 0.06 & 1.8665 &  0.0003 &  0.0003 \\
8    & 17834.62  & 0.03   & 0.05 & 1.8487 &  0.0004 &  0.0004 \\
9    & 19934.09  & 0.01   & 0.04 & 1.8307 &  0.0005 &  0.0005 \\
10   & 22004.52  & 0.00   & 0.03 & 1.8128 &  0.0005 &  0.0005 \\
11   & 24045.86  &-0.01   & 0.03 & 1.7947 &  0.0005 &  0.0006 \\
12   & 26058.03  & 0.00   & 0.05 & 1.7766 &  0.0006 &  0.0006 \\
13   & 28040.94  & 0.03   & 0.07 & 1.7584 &  0.0006 &  0.0006 \\
14   & 29994.52  & 0.08   & 0.10 & 1.7402 &  0.0005 &  0.0006 \\
15   & 31918.65  & 0.13   & 0.14 & 1.7219 &  0.0004 &  0.0005 \\
16   & 33813.23  & 0.17   & 0.17 & 1.7035 &  0.0003 &  0.0004 \\
17   & 35678.10  & 0.17   & 0.16 & 1.6851 & -0.0002 &  0.0002 \\
18   & 37513.11  & 0.08   & 0.08 & 1.6666 & -0.0005 &  0.0000 \\
19   & 39318.06  &-0.19   &-0.11 & 1.6480 & -0.0010 & -0.0003 \\
\midrule
RMSD & 0.00      & 0.18   & 0.08 & 0.0000 &  0.0004 &  0.0004 \\
\midrule\bottomrule
\end{tabular}
\end{table}

\subsection{High-lying vibrational levels for $v$=20-61}
\begin{figure}[H]
    \centering 
    \caption{The difference between MLR$_3$(9) and MLR$_4$(6,8) for all vibrational levels up to the dissociation limit.}
    \includegraphics[width=0.4\textwidth]{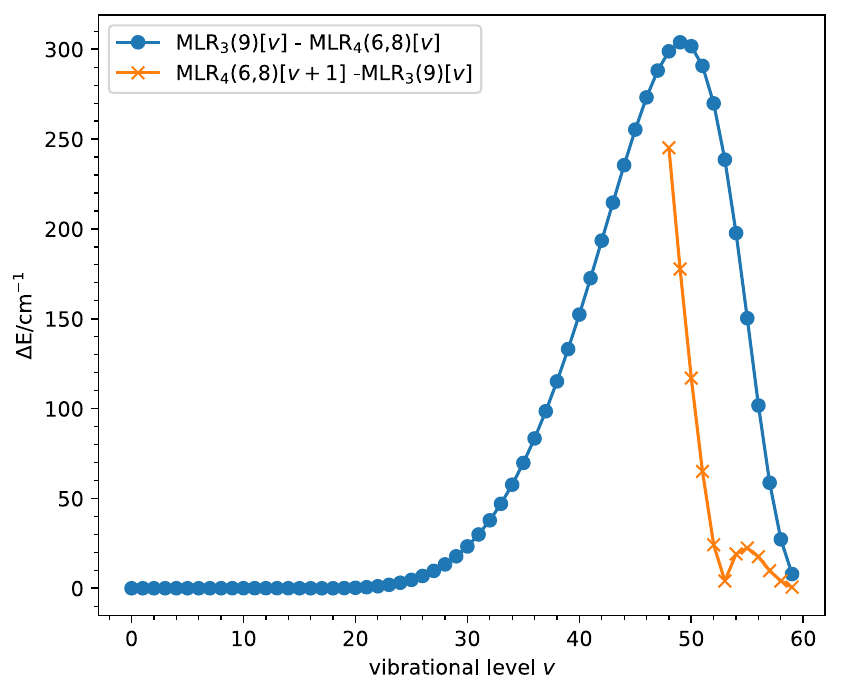}
    \label{fig:N2}
\end{figure}

At dissociation limit, $v_\mathcal{D} \approx 60.8$  MLR$_4$(6,8) $v_\mathcal{D} \approx 59.3$  MLR$_3$(9). These $v_\mathcal{D}$ values suggest that MLR$_4$(6,8) and MLR$_3$(9) support 61 and 59 vibrational levels, respectively.  Differences in low-lying levels are negligible (less than 0.1 cm-1 for $v=$0 to 19), but significant for high-lying levels up to 304 cm-1 at $v=$49 as shown in \Cref{fig:N2}. 

\subsection{LAVA scaling laws across the PEC}
\begin{figure}[H]
    \centering 
    \caption{Power-law scaling trends of Variance vs. Number of parameters.}
    \includegraphics[width=1.0\textwidth]{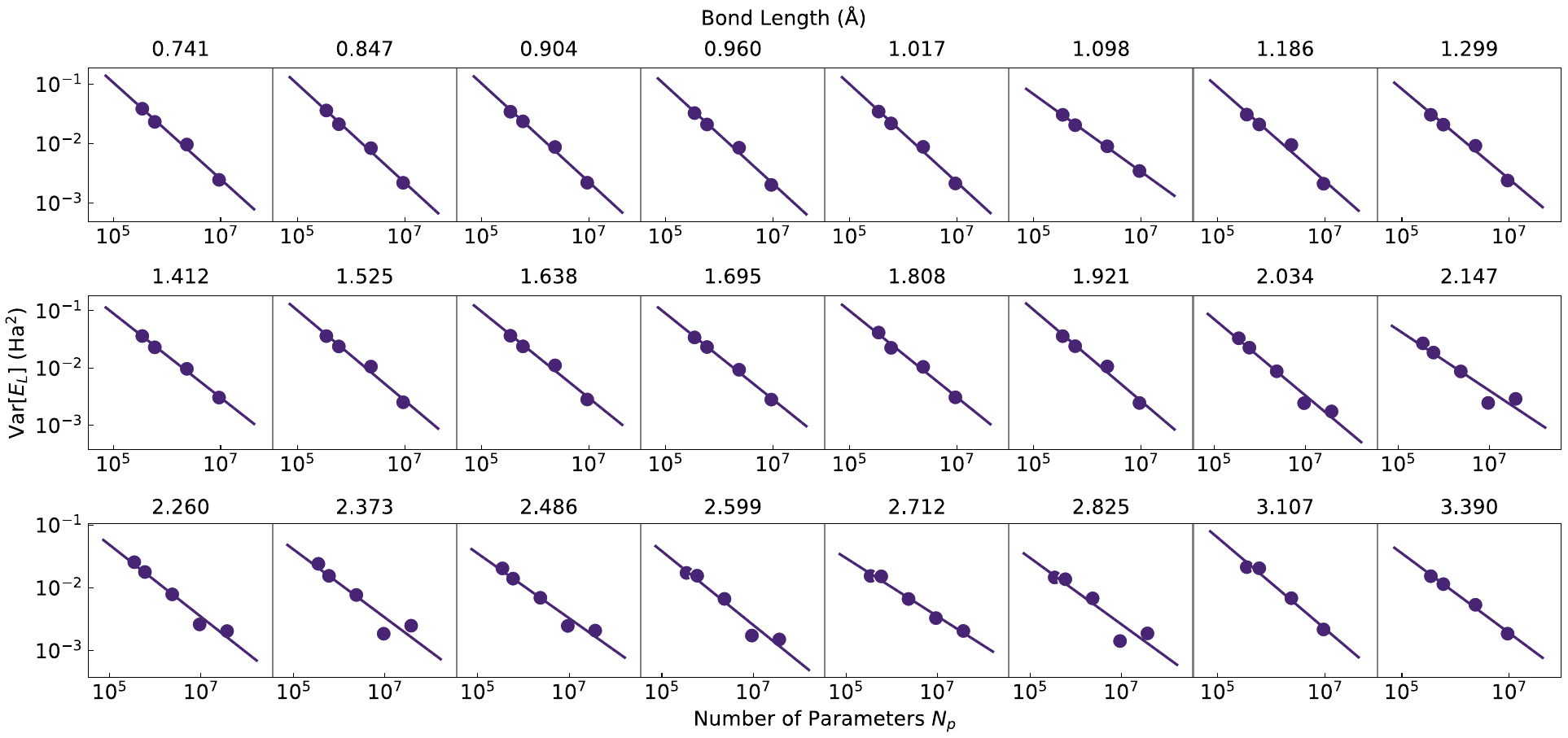}
    \label{fig:scaling_N2_vp}
\end{figure}

\begin{figure}[H]
    \centering 
    \caption{Power-law scaling trends of $E-E_\text{SE}$ vs. Total GPU hours.}
    \includegraphics[width=1.0\textwidth]{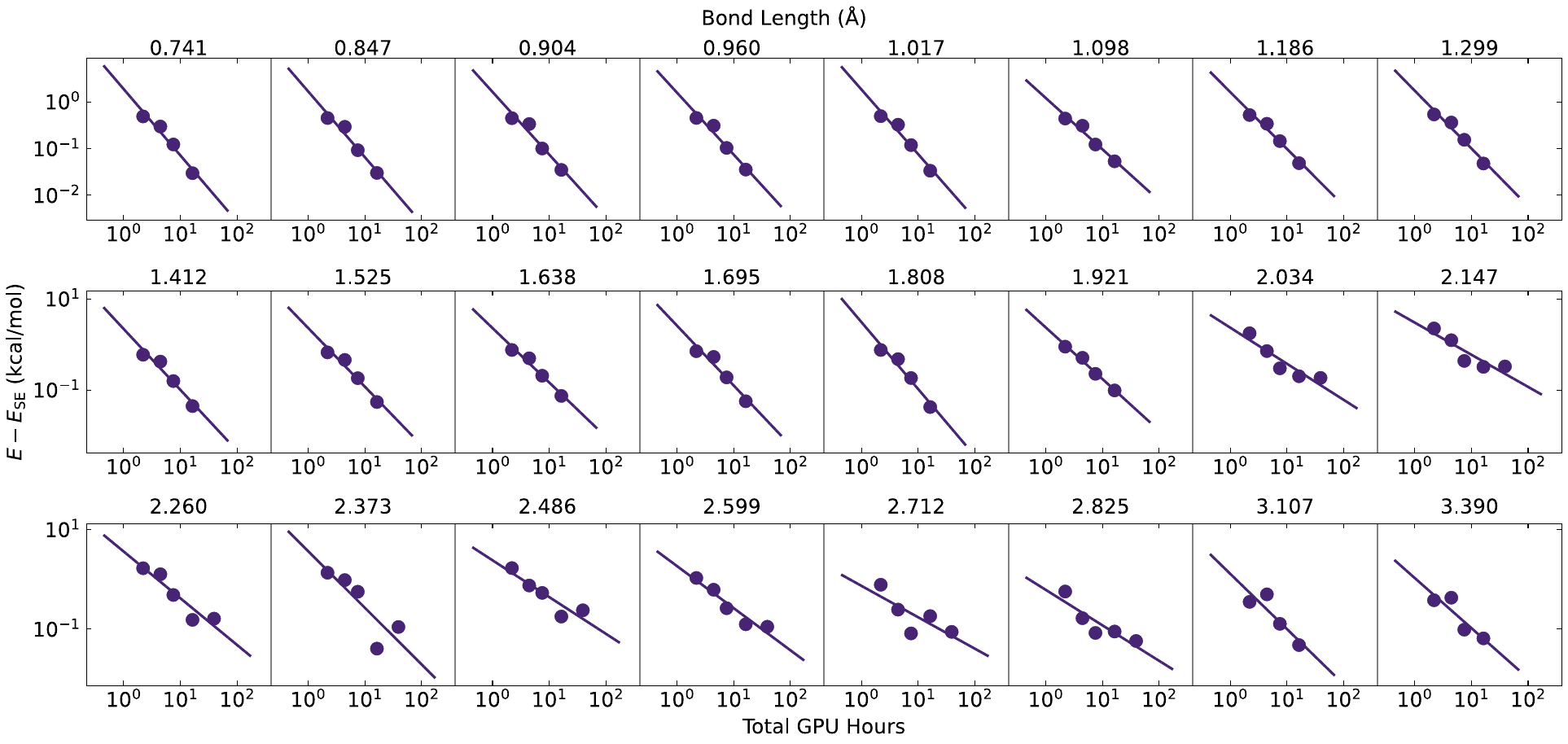}
    \label{fig:scaling_N2_et}
\end{figure}

We tested the scaling laws of LAVA across the PEC using four network configurations: (4,96,2,16), (4,128,4,16), (4,256,4,32), and (4,512,8,128). In \Cref{fig:scaling_N2_vp}, we plot the variance of LAVA against the number of network parameters. The variance consistently decreases to nearly 10$^{-3}$ Ha$^2$ across all bond distances, indicating that LAVA maintains similar accuracy across different bond lengths. \Cref{fig:scaling_N2_et} presents power-law scaling of $E-E_\text{SE}$ against total GPU hours.  
To describe the nearly dissociation regime more accurately, we supplemented the dataset of bond lengths ranging from 2.034 to 2.825 $\mathrm{\AA}$ with results from the (1024,16,256) network. For the PEC shown in the main text Fig. 3b, we adopted the LAVA results from the largest network employed for each specific bond length since the number of data points is not sufficient for reliable extrapolation. But still, in the main text Fig. 2c, we roughly evaluate efficiency scaling across the PEC based on power-law scaling trends between $E-E_{\mathrm{SE}}$ and time costs since we only investigate a qualitative trend of scaling exponent instead of accurate quantitative results of efficiency scaling order.

\section{O$_3$ reaction barrier} \label{sec:si o3}

\subsection{Optimized geometries from various theoretical methods}
As shown in \Cref{tab:O3 geom}, the geometries of open minimum (OM), ring minimum (RM), and transition state (TS) calculated at different levels of theory differ significantly from one another.  Starting from XMS-CASPT2-optimized geometries by perturbing angles and bond lengths, we performed a rough scan of the O$_3$ potential energy surface using (4,256,4,16) LAVA network configuration. Among the methods tested, only LAVA perfectly reproduces the experimental OM geometry. XMS-CASPT2 shows much better agreement with LAVA than CASSCF does: CASSCF overestimates the bond lengths of OM, RM, and TS, as well as the bond angle of TS; XMS-CASPT2 
 overestimates the bond angle of OM and underestimates the bond angle of TS.

\subsection{Energy landscape based on different PESs}
Different methods show good consistency in the relative energy of RM-OM but diverge significantly in the relative energy of TS-RM. We identified three sources of these discrepancies.

First, some studies searched for the transition state geometry under C\textsubscript{2v} symmetry\cite{Theis2016}, others used low  C\textsubscript{s} symmetry\cite{Varga2017}, and still others imposed no spatial symmetry restrictions. Given that $^1\mathrm{A}_1$ and $^1\mathrm{A}_2$ PESs cross near the TS, studies enforcing C\textsubscript{2v}/C\textsubscript{s} symmetry found the TS on $^1\mathrm{A}_1$/$^1\mathrm{A}^\prime$ PES, whereas those without spatial symmetry constraints found the energetically lower cross point between  $^1\mathrm{A}_1$ and $^1\mathrm{A}_2$ surfaces (CP1 in \Cref{tab:O3 geom}) as the TS. 

Second, the quality of the potential energy surfaces (PESs)—particularly their curvature—is strongly dependent on the theoretical method.  CASSCF PESs differ significantly from CASPT2 PESs, whereas CASPT2 PESs show good agreement with LAVA for the critical structures listed in \Cref{tab:O3 geom}. For CASSCF and MRCISD+Q, the structures of TS and CP1  are different by about 4 $^{\circ}$, \cite{Chen2011} while those from XMS-CASPT2 and MMVMC are nearly identical.

Third, even for the same spin state and geometry, significant discrepancies persist across different theoretical methods. In \Cref{tab:O3 barrier singlet}, variations of $\sim$ 0.6 eV arise in the TS-OM relative energy among methods when using the CASSCF geometry and singlet state. \citet{Ghanem2020} calculated these energy gaps using CCSD(T), CCSDT, CCSDT(Q), CCSDTQ, CCSDTQ(P), , with results summarized in \Cref{tab:O3 barrier singlet}. The RM-OM energy converges at the CCSDT(Q) level, but the TS-RM energy continues oscillating strongly even at the CCSDTQ(P) level. In addition, enlarging the basis set from VDZ to VQZ increases the barrier height,  degrading the CC results.  These abnormal behaviors suggest that single-reference coupled cluster theory struggles to accurately describe the TS. Different multireference methods also show a large discrepancy in predicting the TS - RM barrier, as shown in the main text Fig. 4b.

In short, the example of ozone demonstrates LAVA's capability to find the ground state of molecules with complex electronic structures. It is a black-box and systematically improvable. It doesn't require users' chemical intuition to choose active space, nor need any presumption on the ground state characters (such as singlet or triplet, open-shell singlet or closed-shell singlet).

\begin{table}[htbp]
\centering
\caption{Critical structures optimized at different levels of theory.}
\label{tab:O3 geom}
\sisetup{separate-uncertainty, table-format = -1.6}
\begin{threeparttable}
\begin{tabular}{p{8cm} b{2.25cm} b{2.25cm}}
\toprule
\midrule
 \multirow{2}{*}{\makecell{Theory}} & \multicolumn{2}{c}{\textbf{OM}} \\
\cmidrule(lr){2-3}
 & $R_\text{OO}$ & $\angle$OOO  \\
\midrule
CASSCF \cite{Theis2016}               & 1.292        & 116.5    \\
B3LYP\cite{Chen2011}                  & 1.256        & 118.2    \\
XMS-CASPT2\cite{Varga2017}\tnote{a}   & 1.268        & 119.4    \\
LAVA       \tnote{b}                  & 1.278        & 116.8    \\
Experiment\cite{Herzberg1966}         & 1.278        & 116.8    \\
\midrule

\addlinespace[4pt]
 \multirow{2}{*}{\makecell{Theory}} & \multicolumn{2}{c}{\textbf{RM}} \\
\cmidrule(lr){2-3}
& $R_\text{OO}$ & $\angle$OOO\\
\midrule
CASSCF \cite{Theis2016}               & 1.466        &  60.0   \\
B3LYP\cite{Chen2011}                  & 1.432        &  60.0   \\ 
XMS-CASPT2\cite{Varga2017}\tnote{a}   & 1.425        &  60.0   \\
LAVA   \tnote{b}                                & 1.426        &  59.9   \\
\midrule

\addlinespace[4pt]
 \multirow{2}{*}{\makecell{Theory}}  & \multicolumn{2}{c}{\textbf{TS}} \\
\cmidrule(lr){2-3}
 & $R_\text{OO}$ & $\angle$OOO\\
\midrule
CASSCF \cite{Theis2016}               & 1.424        &  84.1    \\
B3LYP\cite{Chen2011}                  & 1.381        &  77.9    \\
XMS-CASPT2\cite{Varga2017}\tnote{a}   & 1.390        &  81.0    \\
LAVA    \tnote{b}                     & 1.390        &  84.0    \\
\midrule

\addlinespace[4pt]
\multirow{2}{*}{\makecell{Theory}} & \multicolumn{2}{c}{C\textbf{ross point}\tnote{c}} \\
\cmidrule(lr){2-3}
& $R_\text{OO}$ & $\angle$OOO\\
\midrule
CASSCF(18,12)/aug-cc-pVTZ \cite{Chen2011}    & 1.416        &  80.3    \\
MRCISD+Q(18,12)/aug-cc-pVTZ\cite{Chen2011}   & 1.416        &  79.7    \\
XMS-CASPT2\cite{Varga2017}            & 1.391        &  80.9    \\
LAVA\tnote{b}                         & 1.394        &  80.9    \\
\midrule
\bottomrule
\end{tabular}
    \begin{tablenotes}
    \item[a] XMS-CASPT2 energies are corrected by  CCSDT(Q) energies by dynamical scaled external correlation (DSEC) at critical structures.\cite{Varga2017} 
    \item[b] We performed a rough scan of the PES using (4,256,4,16) LAVA network configuration.
    \item[c] Cross point between $^1\mathrm{A}_1$ and $^1\mathrm{A}_2$ surfaces that on the RM side. Denoted as CP1 in \citet{Chen2011}.
    \end{tablenotes}
    \end{threeparttable}
\end{table}

\begin{table}[htbp]
\centering
\caption{Geometry effect to relative energies for $^1\mathrm{A}^\prime$ state. Relative energies of TS with respect to OM, RM with respect to OM, and TS with respect to RM. Energies are in eV.}
\label{tab:O3 barrier singlet}

\begin{threeparttable}
\sisetup{separate-uncertainty, table-format = -1.6} 
\begin{tabular}{p{8cm} b{1.5cm} b{1.5cm} b{1.5cm}}
\toprule
\midrule
\textbf{Theory} &  \textbf{TS-OM} &  \textbf{RM-OM} &  \textbf{TS-RM}  \\
\midrule
\addlinespace[4pt]
Single-point energy & \multicolumn{3}{c}{CASSCF geometry\cite{Theis2016}}\\
\cmidrule(lr){1-1} \cmidrule(lr){2-4}
CCSD(T)/VDZ \cite{Ghanem2020}         & 2.55         & 1.38    & 1.18    \\
CCSDT/VDZ \cite{Ghanem2020}           & 2.49         & 1.39    & 1.09    \\
CCSDT(Q)/VDZ \cite{Ghanem2020}        & 2.27         & 1.46    & 0.80    \\
CCSDTQ/VDZ \cite{Ghanem2020}          & 2.34         & 1.44    & 0.90    \\
CCSDTQ(P)/VDZ \cite{Ghanem2020}       & 2.29         & 1.45    & 0.84    \\
\cmidrule(lr){1-1} \cmidrule(lr){2-4}
CCSD(T)/VTZ \cite{Ghanem2020}         & 2.66         & 1.23    & 1.43    \\
CCSDT/VTZ \cite{Ghanem2020}           & 2.60         & 1.25    & 1.36    \\
CCSDT(Q)/VTZ \cite{Ghanem2020}        & 2.35         & 1.32    & 1.02    \\
\cmidrule(lr){1-1} \cmidrule(lr){2-4}
CCSD(T)/VQZ \cite{Ghanem2020}         & 2.72         & 1.27    & 1.45    \\
CCSDT/VQZ \cite{Ghanem2020}           & 2.67         & 1.29    & 1.38    \\
CCSDT(Q)/VQZ \cite{Ghanem2020}        & 2.40         & 1.37    & 1.03    \\
\cmidrule(lr){1-1} \cmidrule(lr){2-4}
CASSCF(18,12)/VQZ\cite{Theis2016}    & 2.32         & 1.33    & 0.99    \\
SHCI(18,90)/VTZ \cite{Chien2018}      & 2.41         & 1.30    & 1.11    \\
FCIQMC(18,39)-Tailored Coupled Cluster/VTZ \cite{Vitale2020}   & 2.43         & 1.29    & 1.14    \\
Adaptive shift FCIQMC \cite{Vitale2020}/VQZ& 2.44         & 1.36    & 1.08    \\
CASSCF(18,12)/VQZ                   & 2.37        & 1.32    & 1.05   \\
NEVPT2(18,12)/VQZ                   & 2.47        & 1.27    & 1.20    \\
\midrule

\addlinespace[4pt]
Single-point energy & \multicolumn{3}{c}{XMS-CASPT2 geometry\cite{Varga2017}}   \\
\cmidrule(lr){1-1} \cmidrule(lr){2-4}
XMS-CASPT2 \cite{Varga2017} \tnote{a}       & 2.42       & 1.19    & 1.23    \\
CASSCF(18,12)/VQZ                   & 2.18       & 1.35    & 0.83   \\
NEVPT2(18,12)/VQZ                   & 2.22       & 1.24    & 0.98   \\
LAVA\tnote{b}                      & 2.27       & 1.36    & 0.91    \\     
\midrule

\addlinespace[4pt]
Single-point energy & \multicolumn{3}{c}{LAVA geometry} \\
\cmidrule(lr){1-1} \cmidrule(lr){2-4}
CASSCF(18,12)/VQZ                   & 2.36        & 1.37    & 0.99   \\
NEVPT2(18,12)/VQZ                   & 2.40        & 1.26    & 1.14   \\
LAVA-SE                      & 2.47        & 1.37    & 1.10   \\  
\midrule
\bottomrule
\end{tabular}

\begin{tablenotes}
\item[a] XMS-CASPT2 energies are corrected by  CCSDT(Q) energies by dynamical scaled external correlation (DSEC) at critical structures.\cite{Varga2017} 
\item[b] LAVA results from (4,512,8,64) network configuration.
\end{tablenotes}

\end{threeparttable}
\end{table}

\begin{table}[htbp]
\centering
\caption{Geometry effect on relative energies for triplet. Relative energies of TS with respect to OM, RM with respect to OM, and TS with respect to RM. Energies are in eV.}
\label{tab:O3 barrier triplet}
\sisetup{separate-uncertainty, table-format = -1.6}
\begin{threeparttable}
\begin{tabular}{p{8cm} b{1.5cm} b{1.5cm} b{1.5cm}}
\toprule
\midrule
\textbf{Theory} &  \textbf{TS-OM} &  \textbf{RM-OM} &  \textbf{TS-RM}  \\
\midrule
\addlinespace[4pt]
Single-point energy & \multicolumn{3}{c}{CASSCF geometry\cite{Theis2016}} \\
\cmidrule(lr){1-1} \cmidrule(lr){2-4}
CASSCF(18,12)/VQZ                   & 1.90        & 1.36    & 0.54   \\
NEVPT2(18,12)/VQZ                   & 1.77        & 1.29    & 0.48   \\
LAVA\tnote{a}                       & 1.82        & 1.36    & 0.46   \\
\midrule

\addlinespace[4pt]
Single-point energy & \multicolumn{3}{c}{XMS-CASPT2 geometry\cite{Varga2017}} \\
\cmidrule(lr){1-1} \cmidrule(lr){2-4}
CASSCF(18,12)/VQZ                   & 2.09        & 1.38    & 0.83   \\
NEVPT2(18,12)/VQZ                   & 1.87        & 1.26    & 0.98   \\
LAVA\tnote{a}                       & 1.93        & 1.35    & 0.58   \\     
\midrule

\addlinespace[4pt]
Single-point energy & \multicolumn{3}{c}{LAVA geometry} \\
\cmidrule(lr){1-1} \cmidrule(lr){2-4}
CASSCF(18,12)/VQZ                   & 2.50        & 1.40    & 1.10  \\
NEVPT2(18,12)/VQZ                   & 2.22        & 1.27    & 0.95  \\
LAVA-SE                       & 2.08        & 1.36    & 0.72  \\  
\midrule
\bottomrule
\end{tabular}

\begin{tablenotes}
\item[a] LAVA results from (4,512,8,64) network configuration.
\end{tablenotes}
\end{threeparttable}

\end{table}

\subsection{Half-life estimation}
In classical mechanics, chemical reactions can only occur at energies exceeding the reaction barrier. Classical reaction rates can be estimated using the Arrhenius equation.  Currently, calculating vibrational frequencies with LAVA remains computationally prohibitive. Thus, we adopted the thermochemistry correction based on the geometries and vibrational frequencies (summarized in \Cref{tab:O3 parition function}) at XMS-CASPT2 level by \citet{Varga2017}, given that XMS-CASPT2 shows good agreement with LAVA for critical structures at \Cref{tab:O3 geom}. As for electronic energies, we used the LAVA-SE energies based on LAVA-optimized geometries, as listed in \Cref{tab:O3 parition function}. Half-lives are calculated from the rate constants assuming first-order reactions. Rate constants and half lives from classical mechanics are provided in \Cref{tab:O3 half-life}, labeled as TST (Conventional transition state theory).

\begin{table}[htbp]
\centering
\caption{Parameters to calculate half-life.}
\label{tab:O3 parition function}
\sisetup{separate-uncertainty, table-format = -1.6} 
\begin{tabular}{llll} 
\toprule\midrule
\textbf{Energy component} &  \textbf{OM} &  \textbf{TS} &  \textbf{RM}  \\
\midrule
\addlinespace[4pt]

Vibrational frequencies (cm$^{-1}$)\cite{Varga2017}   & 1130,756,730 & 929,590,1005$i$    & 1259,1259,1121   \\
LAVA-SE electronic energy (Ha)       & -225.4368        & -225.3459    & -225.3864    \\
\midrule\bottomrule
\end{tabular}
\end{table}

\subsection{Oxygen tunneling}
In quantum mechanics, however, chemical reactions can occur at energies below the reaction barrier via quantum-mechanical tunneling (QMT). QMT results in reaction rates significantly faster than predicted by Arrhenius behavior and is characterized by unusually large kinetic isotope effects (KIEs). Tunneling effects are well-known and particularly pronounced in reactions involving hydrogen or proton transfer, as well as those with high barriers.  Additionally, several experiments have observed pronounced heavy-atom tunneling in ring-opening reactions. For example,  \citet{Datta2008} theoretically predicted rapid carbon tunneling in the ring opening of cyclopropylcarbinyl radical, which was later confirmed by experiments \cite{GonzalezJames2010}. Using canonical variational transition-state theory (CVT) with the small-curvature tunneling (SCT) approximation, the theoretical predictions from \citet{GonzalezJames2010} achieved excellent agreement with experimental results. More recently, \citet{Zhou2023} experimentally observed oxygen tunneling in both the oxygen-oxygen bond breaking reaction of cyclic beryllium peroxide to form linear dioxide and ring-closure reactions of beryllium ozonide complexes. The latter reaction~\cite{Zhou2024} bear strong similarities to the ring-opening of cyclic ozone, and \citet{Zhou2024} demonstrated that  QMT dominates the oxygen-oxygen bond breaking in beryllium peroxide.

\citet{Chen2011} quantified QMT of the ring-opening reaction of cyclic ozone using canonical variational theory (CVT). They employed two approaches to evaluate the microcanonical optimized multidimensional tunneling ($\mu$OMT) correction: one at continuous energy levels (the conventional method) and another at quantized reactant states (the QRST method). To isolate the QMT effect, We took the difference between their CVT/$\mu$OMT-QRST and CVT results in Table 3 of ref.\cite{Chen2011} as the QMT effect, and multiplied it onto our TST results as LAVA's predictions of QMT results.   Rate constants and half-lives from QMT are provided in \Cref{tab:O3 half-life}, labeled as TST (transition state theory).

\begin{table}[htbp]
\centering
\caption{Rate constants and half-life derived from LAVA-SE results. }
\label{tab:O3 half-life}
\sisetup{separate-uncertainty, table-format = -1.6}
\begin{threeparttable}
\begin{tabular}{ccccc} 
\toprule
\midrule
\multirow{2}{*}{\textbf{$T$(K)}} &  \multicolumn{2}{c}{\textbf{Rate constants  (s$^{-1}$)}} &  \multicolumn{2}{c}{\textbf{Half-life (s)}}  \\
\cmidrule(lr){2-3} \cmidrule(lr){4-5}
 & TST & CVT/$\mu$OMT-QRST & TST & CVT/$\mu$OMT-QRST  \\

\midrule
100                   &    $5.6\times10^{-36}$   & $1.2\times 10^{-7}$      &   $1.2\times 10^{35}$    & $5.6\times10^{6}$ \\
200                   &    $3.3\times10^{-11}$     &   $5.3\times10^{-4}$    & $2.1\times10^{10}$      & $1.3\times10^{3}$ \\
300                   &    $7.3\times10^{-3}$    & $1.3\times10^0$      &  $9.5\times10^1$     &  $5.5\times10^{-1}$\\
\midrule
\bottomrule
\end{tabular}
\end{threeparttable}

\end{table}

\subsection{LAVA scaling laws for ozone}
\begin{figure}[htp]
    \centering 
    \caption{Power-law scaling trends for ozone.}
    \includegraphics[width=0.55\textwidth]{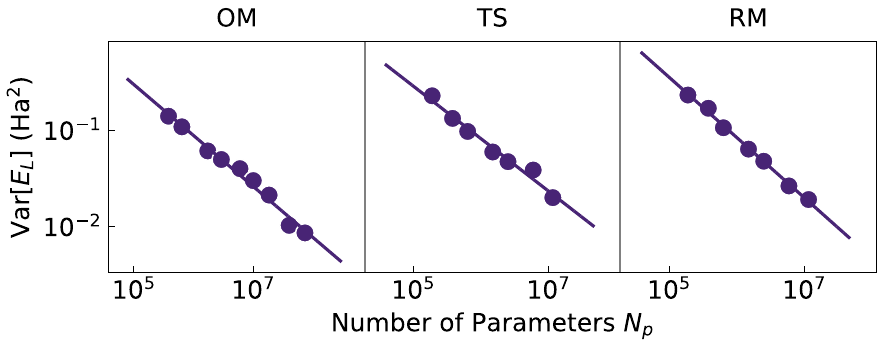}
    \label{fig:fig_ozone_scaling_vp}
\end{figure}

\begin{figure}[htp]
    \centering 
    \caption{Variance-energy extrapolation for ozone.}
    \includegraphics[width=0.9\textwidth]{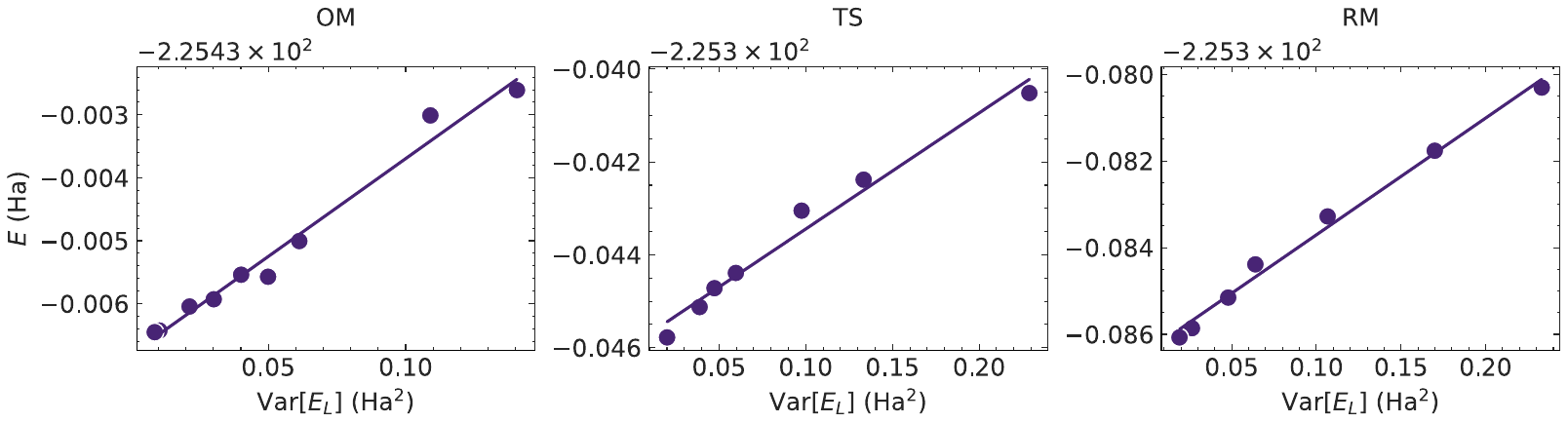}
    \label{fig:fig_ozone_ve}
\end{figure}

\Cref{fig:fig_ozone_scaling_vp} gives scaling laws of the variance of local energy with respect to the number of parameters for different ozone geometries, while \Cref{fig:fig_ozone_ve} shows the variance-energy extrapolation.

\subsection{Improve spin symmetry and spatial symmetry of wavefunctions by scaling up network size}
As illustrated by the XMS-CASPT2 PESs in Fig. 4c, the lowest-energy singlet state switches from $1^1\mathrm{A}^{\prime}$ to $1^1\mathrm{A}^{\prime\prime}$ near the TS, and the state crossing point almost overlaps with the TS (\Cref{tab:O3 geom}). Additionally, the conical intersection between $1^1\mathrm{A}^{\prime}$ and $2^1\mathrm{A}^{\prime}$ states also lies closely to the TS. With the (4,256,4,16) network size, we observed state contamination between $1^1\mathrm{A}^{\prime}$ and $1^1\mathrm{A}^{\prime\prime}$, indicated by the $\sigma_h=-0.62$ in \Cref{table:symmetry_scaling}. Despite the complex electronic structure of the TS, LAVA converges straightforwardly to the correct spatial symmetry with $\sigma_h=-1.00$ as the model size increases to (4,512,8,128).

A similar trend applies to spin symmetry. The CASSCF TS lies near the crossing point between $1^1\mathrm{A}^{\prime}$ and $1^3\mathrm{A}^{\prime\prime}$ states. A singlet state should have an expectation value of the total angular momentum operator $\langle\hat{S}^2\rangle$ that equals 0, while a triplet should have a value of 2. The (4,256,4,16) network gives an $\langle\hat{S}^2\rangle$ value of 1.30, suggesting severe spin contamination in the LAVA wavefunction (\Cref{table:symmetry_scaling}). Scaling up the model to (4,512,8,128) —without imposing symmetry constraints or a penalty loss term\cite{li_spin-symmetry-enforced_2024}—ultimately yields a nearly pure triplet state with $\langle\hat{S}^2\rangle$ value of 1.85.

\begin{table}[htbp]
\centering
\caption{Expectation values of spatial and spin symmetries as the network scales up. }
\sisetup{separate-uncertainty, table-format = -1.6}
\begin{threeparttable}
\begin{tabular}{cS[table-format=5.7]S[table-format=5.7]S[table-format=5.7]S[table-format=5.7]} 
\toprule
\midrule
\multirow{2}{*}{\makecell{Network\\ configuration}} & \multicolumn{2}{c}{Spatial symmetry at XMS-CASPT2 TS}  &  \multicolumn{2}{c}{Spin symmetry at CASSCF TS}  \\
\cmidrule(lr){2-3} \cmidrule(lr){4-5} 

  & $\sigma_h$ & \text{Energy\ (Ha)}  & \multicolumn{1}{c}{$\langle \hat{S}^2\rangle$} & \text{Energy\ (Ha)} \\

\midrule
(4,256,4,16)    &   -0.62    & -225.3523  &  1.30     &  -225.3664   \\
(4,512,8,128)   &   -1.00     & -225.3572     &    1.85    &  -225.3725  \\
\midrule\bottomrule
\end{tabular}
\vspace{-0.3cm}
\end{threeparttable}
\label{table:symmetry_scaling}
\end{table}

\section{Dipole moments and TAE for multireference molecules}  \label{sec: si dipole}

\subsection{Multireference diagnostics}

We use coupled cluster $\mathcal{T}$1\cite{Lee2009}, $\mathcal{D}$1\cite{Janssen1998}, and \%TAE$_e$[(T)] diagnostics \cite{Karton2006} to quantify the multireferential character of tested molecules. \citet{Lee2009} used the criteria of $\mathcal{T}1 > 0.02$ and  \citet{Janssen1998} used $\mathcal{D}1 > 0.05$ to indicate non-negligible multireferential characters.

\citet{Karton2006} quantified the multireferential character using \%$\text{TAE}_e$[(T)]:

\begin{equation}
\%\text{TAE}_e[(\text{T})]=100 \times \frac{\text{TAE}_e[\text{CCSD(T)}]-\text{TAE}_e[\text{CCSD}]}{\text{TAE}_e[\text{CCSD(T)}]}
\end{equation}

A \%TAE$_e$[(T)] value below 2\% indicates systems are dominated by dynamical correlation; 2–5\% indicates mild nondynamical correlation; 5–10\%indicates  moderate nondynamical correlation; and values in excess of 10\% indicate severe nondynamical correlation. Multireference diagnostics values taken from \citet{karton_w4-11_2011} are listed in \Cref{tab:diagnostics}.

\begin{table}[htbp]
\centering
\caption{Multireference diagnostics for molecules in  Fig. 5a.  }
\label{tab:diagnostics}
\begin{threeparttable}
\sisetup{separate-uncertainty, table-format = -1.6} 
\begin{tabular}{l S[table-format=1.8] S[table-format=1.8] S[table-format=1.8]}  
\toprule
\midrule
\textbf{Molecule} &  \textbf{$\mathcal{T}$1 diagnostic} &  \textbf{$\mathcal{D}$1 diagnostic} &  \textbf{\%TAE$_e$[(T)]}  \\
\midrule
CH$_3$COOH    & 0.02        & 0.05      & 2.0           \\
HN$_3$        & 0.02        & 0.05      & 5.6           \\
NO            & 0.02        & 0.05      & 6.2           \\
F$_2$O$_2$    & 0.03        & 0.09      & 16.9          \\
O$_3$         & 0.03        & 0.08      & 17.4          \\
\midrule\bottomrule
\end{tabular}
\end{threeparttable}
\end{table}

\subsection{Dipole moment}

\citet{Hait2018} recommended combining the SCF component at the aug-cc-pCVQZ level with correlated component—estimated via extrapolation from aug-cc-pCVQZ and aug-cc-pCVTZ basis sets—to obtain a practically useful estimate of $\mu$ when expensive aug-cc-pCV5Z calculations are not affordable. For HN$_3$ and O$_3$, we adopt the CCSD(T)/CBS results from \citet{Hait2018}. For molecules not included in their work, we use the largest available basis sets from the Basis Set Exchange website\cite{Pritchard2019} as the CBS limit of HF calculations, namely cc-pV6Z, aug-cc-pV6Z, and aug-cc-pCV5Z. For CCSD and CCSD(T), we use V{Q,5}Z and V{D,T}Z basis set pairs, respectively,  to estimate the CBS limit for the correlated component in \Cref{tab:foof dipole}. 

Owing to the slow basis set convergence of post-HF methods, \citet{Hait2018}  employed the two-point extrapolation scheme from \citet{Halkier1999dipole}, which reproduces accurate (approximately 0.2\% error) for small molecules—yielding results comparable to quintuple-zeta quality from triple- and quadruple-zeta calculations. 
\begin{equation}
    \mu^{\rm{corr}}_{\infty} = \frac{n^3 \mu^{\rm{corr}}_n - m^3 \mu^{\rm{corr}}_m}{n^3 - m^3} ,
\end{equation}
where $n$ and $m$ is the $\zeta$ cardinality for basis set.

\begin{table}[htbp]
\centering
\caption{Coupled cluster calculations of the dipole moment  of $\text{F}_2\text{O}_2$, reported in Debyes.}
\label{tab:foof dipole}
\begin{threeparttable}
\sisetup{separate-uncertainty, table-format = -1.6} 
\begin{tabular}{ p{3cm} p{2.5cm} p{2.5cm} p{2.5cm}}  
\toprule
\midrule
\textbf{Theory} &  \textbf{Hartree-Fock}\tnote{a} &  \textbf{CCSD}\tnote{a} &  \textbf{CCSD(T)}\tnote{b}  \\
\midrule
cc-pVDZ        & 1.37        & 1.19      & 1.20         \\
cc-pVTZ        & 1.49        & 1.29      & 1.28         \\
cc-pVQZ        & 1.51        & 1.33      & OOM\tnote{c} \\
cc-pV5Z        & 1.52        & 1.36      & OOM          \\
cc-pV6Z        & 1.52        & OOM       & OOM          \\
\addlinespace[0.5ex]  
\hdashline          
\addlinespace[0.5ex]
CBS limit & 1.52            & 1.37       & 1.30        \\
(3,4) &             & 1.35      & N/A        \\ 
\midrule
aug-cc-pVDZ    & 1.52        & 1.41      & 1.40 (with ccecp) \tnote{d}       \\
aug-cc-pVTZ    & 1.52        & 1.36      & 1.38 (with ccecp)        \\
aug-cc-pVQZ    & 1.52        & 1.36      & OOM          \\
aug-cc-pV5Z    & 1.51        & 1.36      & OOM          \\
aug-cc-pV6Z    & 1.51        & OOM       & OOM          \\
\addlinespace[0.5ex]  
\hdashline         
\addlinespace[0.5ex]
CBS limit &   1.51          & 1.36      & 1.37          \\ 
\midrule
aug-cc-pCVDZ    & 1.52       & 1.41      & 1.44         \\
aug-cc-pCVTZ    & 1.52       & 1.36      & OOM          \\
aug-cc-pCVQZ    & 1.52       & 1.36      & OOM          \\
aug-cc-pCV5Z    & 1.51       & 1.36      & OOM          \\
\addlinespace[0.5ex]  
\hdashline        
\addlinespace[0.5ex]
CBS limit  &    1.51         & 1.36      & N/A          \\ 
\midrule
\addlinespace[2pt]
\text{LAVA }    & \multicolumn{3}{c}{1.40 $\pm$ 0.01 }  \\
\midrule
\addlinespace[2pt]
\textbf{Experiment}     & \multicolumn{3}{c}{1.44 $\pm$ 0.04}\cite{Jackson1962} \\
\addlinespace[2pt]
\midrule
\bottomrule
\end{tabular}
\begin{tablenotes}
\item[a] Calculations performed on ByteQC \cite{Guo2025} with GPU-accelerated HF and CCSD modules. The GPU-accelerated CCSD(T) module is not available.
\item[b] Calculations performed on PySCF v2.9.0.\cite{Sun2020}
\item[c] OOM = out of memory on Intel(R) Xeon(R) Platinum 8336C CPU @ 2.30GHz, 980 GB memory.
\item[d] All-electron calculations ran out of memory. 
\end{tablenotes}

\end{threeparttable}
\end{table}

Dipole moments calculated by post-HF methods are highly sensitive to the choice of basis set. The inclusion of augmented functions (denoted as "aug-") is critical for obtaining reliable dipole moments with relatively small basis sets. However, the addition of diffuse functions rapidly renders coupled-cluster calculations intractable due to their high computational scaling.   For $\text{F}_2\text{O}_2$ molecules, both the diffuse and high-angular-momentum basis functions are critical to obtain an accurate dipole moment comparable to experimental results. In \Cref{tab:foof dipole}, the CCSD/CBS dipole moment remains 0.04 Debye from the experimental value. While CCSD(T)/CBS holds promise for reproducing the experimental results. However, its prohibitive $\mathcal{O}(N^7)$ computational scaling renders all-electron CCSD(T) calculations with the aug-cc-pVTZ and aug-cc-pCVTZ basis sets unfeasible.

\begin{figure}[htp]
    \centering 
    \caption{F$_2$O$_2$ dipole results of different networks.}
    \includegraphics[width=.47\textwidth]{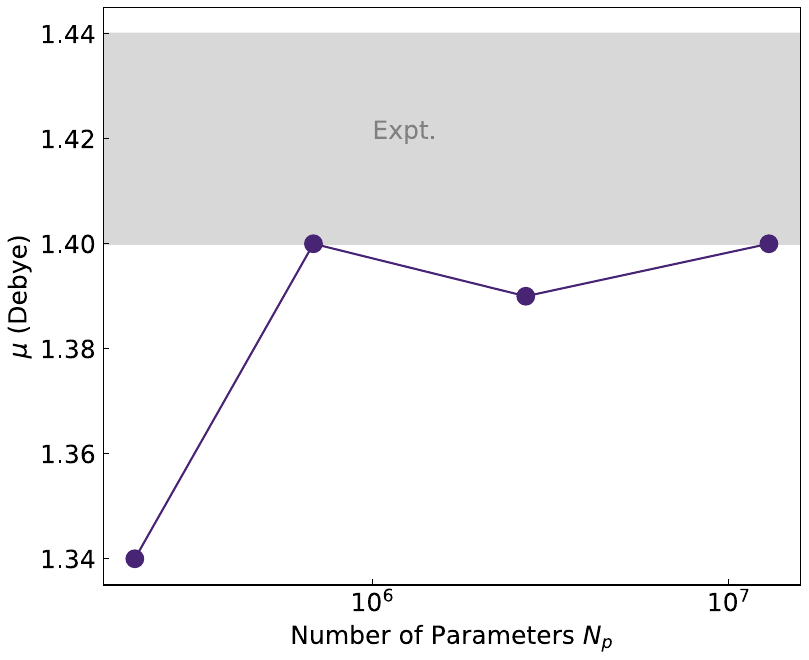}
    \label{fig:foof_dipole_scaling}
\end{figure}

In contrast, as shown in Fig. 5a, LAVA yields highly accurate dipole moments, highlighting its advantage of delivering FCI/CBS-quality results, benefiting from its first-quantized ansatz.  \Cref{fig:foof_dipole_scaling} further demonstrates that the dipole moment results of F$_2$O$_2$ rapidly converge to the experimental range as the network scales up. 

\bibliography{si-bibliography}